\documentclass[aps,prd,floatfix,column,superscriptaddress,preprintnumbers,nofootinbib,preprint]{revtex4}
\usepackage{graphicx}

\usepackage{ulem}
\usepackage{color}
\definecolor{My_red}        {cmyk}{0.00,1.00,1.00,0.20}

\newcommand{\bmat}{\left(\begin{array}}
\newcommand{\emat}{\end{array}\right)}
\newcommand{\beq}{\begin{equation}}
\newcommand{\eeq}{\end{equation}}

\def\ra{\rightarrow}

\def\Ld{\Lambda}
\def\ld{\lambda}
\def\f{\frac}

\def\bwt{\begin{widetext}}
\def\ewt{\end{widetext}}
\def\be{\begin{equation}}
\def\ee{\end{equation}}
\def\bea{\begin{align}}
\def\eea{\end{align}}
\def\bean{\begin{align*}}
\def\eean{\end{align*}}
\def\bary{\begin{array}}
\def\eary{\end{array}}
\def\bit{\begin{itemize}}
\def\eit{\end{itemize}}

\def\ra{\rightarrow}

\def\Ld{\Lambda}

\def\ld{\lambda}

\def\su5u1{SU(5) \times U(1)}
\def\fsu5u1{SU(5) \times U(1)_X }
\def\so10{SO(10)}
\def\sq20{SO(10) \times SO(10)}

\def\ra{\rightarrow}

\def\Ld{\Lambda}
\def\ld{\lambda}
\def\f{\frac}

\def\L{\left(}
\def\R{\right)}
\def\bwt{\begin{widetext}}
\def\ewt{\end{widetext}}
\def\be{\begin{equation}}
\def\ee{\end{equation}}
\def\bea{\begin{align}}
\def\eea{\end{align}}
\def\bean{\begin{align*}}
\def\eean{\end{align*}}
\def\bary{\begin{array}}
\def\eary{\end{array}}
\def\bit{\begin{itemize}}
\def\eit{\end{itemize}}

\def\ra{\rightarrow}

\def\Ld{\Lambda}

\def\ld{\lambda}

\def\su5u1{SU(5) \times U(1)}
\def\fsu5u1{SU(5) \times U(1)_X }
\def\so10{SO(10)}
\def\sq20{SO(10) \times SO(10)}

\usepackage[centertags]{amsmath}
\usepackage{amssymb}

\begin{document}

\title{Accidental Dark Matter: \\ 
Case in the Scale Invariant Local $B-L$ Model }

%--------------------------------------------------------------------------------------------------
\author{Jun Guo}
\email[E-mail: ]{hustgj@itp.ac.cn}
 \affiliation{State Key Laboratory of
Theoretical Physics and Kavli Institute for Theoretical Physics
China (KITPC), Institute of Theoretical Physics, Chinese Academy of
Sciences, Beijing 100190, P. R. China}

\author{Zhaofeng Kang}
\email[E-mail: ]{zhaofengkang@gmail.com}
\affiliation{School of Physics, Korea Institute for Advanced Study,
Seoul 130-722, Korea}

\author{P. Ko}
\email[E-mail: ]{pko@kias.re.kr}
\affiliation{School of Physics, Korea Institute for Advanced Study,
Seoul 130-722, Korea}

\author{Yuta Orikasa}
\email[E-mail: ]{orikasa@kias.re.kr}
\affiliation{School of Physics, Korea Institute for Advanced Study,
Seoul 130-722, Korea}
\affiliation{Department of Physics and Astronomy,
Seoul National University, Seoul 151-742, Korea}

\date{\today}

\begin{abstract}
We explore the idea of accidental dark matter (aDM) stability in the scale invariant local $U(1)_{B-L}$ model, 
which is a theory for neutrino and at the same time radiatively breaks scale invariance via quantum 
mechanical  dynamics in the $U(1)_{B-L}$ sector. A real singlet scalar can be accidental DM with an accidental $Z_2$, by virtue of both extended symmetries. A $U(1)_{B-L}$ charged complex scalar can also be a viable accidental DM due to an accidental (or remanent) $Z_3$. They can reproduce correct relic density via the annihilations through the conventional Higgs portal or dark Higgs portal. The dark Higgs portal scenario is in tension with the LHC bound on $Z_{B-L}$, and only heavy DM of a few TeVs can have correct relic density. In particular,  DM may trigger CSI spontaneous breaking. The situation is  relaxed significantly in the $Z_3$ case due to the effective semi-annihilation mode and then light DM can be accommodated easily. In addition, the $Z_3$ model can accommodate the GeV scale $\gamma-$ray excess from the galactic center via semi-annihilation into pseudo Goldstone boson (PGSB). The best fit is achieved at a DM about 52 GeV, with annihilation cross section consistent with the thermal relic density. The invisible Higgs branching ratio is negligible because the Higgs portal quartic coupling is very small $\lambda_{h\phi} \lesssim 10^{-3}$.

\end{abstract}


%\preprint{MIFP-10-nn}

\maketitle

%---------------------------------------------------------------------------------------------------------
\section{Introduction and motivation}
%---------------------------------------------------------------------------------------------------------

Discovery of the standard model (SM) Higgs-like scalar boson at the LHC inspires new ideas to understand   
the following two basic questions about this special member of SM: How does it develop vacuum expected 
value (VEV), namely break the electroweak (EW) symmetries? And why does its VEV keep as low as 
100 GeV  in the presence of quantum corrections? 
Extending the spacetime symmetries of the SM by including classical scale invariance (CSI) potentially 
provides answers to these two questions simultaneously. This symmetry implies that the cut-off scale used 
in the cut-off regularization turns out to be a tool instead of a physical scale, and consequently the hierarchy 
problem may not be a physical problem~\cite{Bardeen}. 
Moreover, the CSI is anomalous and can be broken by undergoing dimensional transmutation: 
via Coleman-Weinberg (CW) mechanism in the perturbative region ~\cite{CW,hiddenCW,Iso:2009nw}, 
or via confining dynamics in the strong coupling region ~\cite{strong}. 
It opens a chance to understand the origin of EW scale from pure quantum mechanical dynamics. 
 In order to maintain perturbativitiy of the theory up to high energy scale, it is favored to implement the CW mechanism in a hidden sector and then transfer the resulting scale to the SM sector~\cite{hiddenCW}.  Otherwise, one has to turn to a special structure of dynamics~\cite{RGE}.  

Although not necessary, the hidden sector is supposed to contain a gauge group in order to 
adopt  the original CW mechanism. 
It seems that the local $U(1)_{B-L}$ gauge symmetry is a very good case~\cite{Iso:2009nw}. 
Actually, we do have another strong motivation to extend the SM gauge group by introducing $U(1)_{B-L}$, 
i.e., to understand why the observed neutrinos are massive but very light. 
As is well known, incorporating right-handed neutrinos (RHNs) can render active neutrinos massive 
through the elegant canonical seesaw mechanism. In the framework of CSI, introducing extra scalar 
singlets developing VEVs can realize the seesaw mechanism maintaining classical 
scale invariance. Those singlets play an important role in CSI spontaneously breaking~\cite{RHN:SI,Kang:2014cia} 
\footnote{Scale invariant models that radiatively generate neutrino masses are also studied~\cite{Lindner:2014oea}.}. Nevertheless, why we need such RHNs and how many of them we need may have more 
profound physical reasons. As is well known, the SM fermions are not  arbitrary;  they are deliberately 
arranged in order to fulfill gauge anomaly cancelations.
Similarly, the RHNs, with total number three, gain that kind of legitimacy in the local $U(1)_{B-L}$ models. 
Therefore, the scale invariant $B-L$ model provides a good example to address hierarchy problem as well 
as phenomenological drawbacks of SM in the neutrino sector which strongly hint for new physics.

In this paper we address another motivation for physics beyond the SM (BSM), namely lack of a viable 
dark matter (DM) candidate, in the scale invariant $U(1)_{B-L}$ extension  of the SM. 
This topic has been discussed by quite a few groups already, e.g., Refs.~\cite{Carone:2013wla,common,Guo:2014bha,Kang:2014cia}.   In this paper,  with the extended spacetime 
and gauge symmetries at hand,  we attempt to understand one of  the basic questions about 
DM, namely, why is it stable?  The basic idea to address this question was proposed 
in Ref.~\cite{Guo:2014bha,Kang:2014cia}, i.e., DM could
be accidentally stable due to the extended spacetime symmetries and field content, rather 
than a protecting {\it  ad hoc }symmetry imposed by hand. This idea of accidental DM (aDM)  
is motivated by the accidental stability of proton in SM. 

In the context of $U(1)_{B-L}$ extension of the SM,  it is found that two non-trivial scenarios 
can be realized: 
\begin{enumerate}
\item  A real singlet scalar is an accidental DM by an accidental $Z_2$, by virtue of two  
extended symmetries  (classical scale symmetry and the local $U(1)_{B-L}$ gauge symmetry); 
\item A complex scalar charged under $U(1)_{B-L}$ can also be a viable accidental DM due to an accidental 
(or remanent ~\cite{Ko:2014nha}) $Z_3$, which essentially is by virtue of the local $U(1)_{B-L}$ 
but CSI helps to negate other possibilities.
\end{enumerate}
Both scenarios can produce correct thermal relic DM density via the interactions through the Higgs 
portal to dark Higgs field.  But the latter case is in tension with the LHC bound on the $Z_{B-L}$ gauge 
boson from the  Drell-Yan process, and only heavy DM of a few TeVs can have correct relic density.  
Such a heavy scalar DM may lead to an interesting phenomena,  i.e., 
spontaneous breaking of CSI could be triggered by DM  rather than by $Z_{B-L}$. 
The $Z_3$ case is distinguishable from the $Z_2$ case, since it possesses the characteristic 
semi-annihilation channel, which can relax the tension encountered in the $Z_2$ dark model with Higgs 
portal scenario. Moreover, a light DM can be accommodated  easily in the $Z_3$ case, as first pointed 
out in Ref.~\cite{Ko:2014nha}.   For example, the GeV scale $\gamma-$ray excess from the galaxy center can be interpreted by a light DM with mass around 50 GeV and semi-annihilating into a PGSB plus DM.

This paper is organized as follows. 
In Section II, we study symmetry breaking mechanism in the scale invariant $U(1)_{B-L}$ extension 
of  the SM, and its mediation to the SM sector. The feature of the Higgs spectra is also discussed. 
In Section III, we explore accidental DM models with $Z_2$ and $Z_3$ symmetries and phenomenology 
therein. Section IV contains the discussion and conclusion. 
Some supplementary materials are casted in the Appendices.

%---------------------------------------------------------------------------------------------------------
\section{From $B-L$ to electroweak symmetry breaking}
%---------------------------------------------------------------------------------------------------------

The minimal classically scale invariant $U(1)_{B-L}$ model (MSIBL) can naturally solve the 
gauge hierarchy problem, by breaking the electroweak (EW) $SU(2)_L\times U(1)_Y$ 
symmetry without introducing the  bare mass term for the Higgs doublet. By introducing an extra 
$U(1)_{B-L}$  symmetry and a scalar field $\Phi$ which is a SM singlet but carries $U(1)_{B-L}$ 
charge 2, we can break the local $U(1)_{B-L}$ symmetry through the Coleman-Weinberg 
(CW) mechanism~\cite{CW}. Besides, the right-handed neutrinos (RHNs) can acquire 
Majorana masses and realize the standard seesaw mechanism as usual after spontanesous 
breaking of CSI. Then, EWSB can be induced by the Higgs portal   ($\sim H^\dagger H \Phi^\dagger \Phi$)
interaction between the SM Higgs field and a new scalar $\Phi$~\cite{hiddenCW,Iso:2009nw}. 
Let us discuss the details of each aspect in the following, including the current collider bounds  
on the new particles. We will focuses on the $Z_{B-L}$ gauge boson, which will be highly 
relevant to the DM phenomenologies.

%---------------------------------------------------------------------------------------------------------
\subsection{Symmetry breaking and Higgs spectra}\label{SB:spectra}
%---------------------------------------------------------------------------------------------------------

The MSIBL has two types of new fields, a singlet scalar $\Phi$ to break $U(1)_{B-L}$ gauge symmetry 
and three generations of right-handed neutrinos (RHNs) to cancel gauge anomalies. 
The presence of RHNs naturally explains why the SM neutrinos are massive.  
While $U(1)_{B-L}$ breaking naturally furnishes the Majorana mass origins of the RHNs, 
it moreover fixes the $U(1)_{B-L}$ charge of the field $\Phi$ to be $+2$. 
With these new particles, the Lagrangian reads
\begin{align} 
{\cal L}= {\cal L}_{\rm SM} - V(H,\Phi)  -  
\left(\f{1}{2}\ld_{N,i}\Phi \bar N_i^c N_i+Y_{N,ij}\bar \ell H^\dagger N+h.c.\R
\end{align}
We are working in a basis where the mass matrix for the RHNs are diagonal after 
$U(1)_{B-L}$  spontaneously breaking (BLSB). The classical scalar potential with CSI 
takes the following generic form: 
\begin{align}\label{VCW}
V(H,\Phi)=\f{\lambda_{h}}{2}|H|^4 -\ld_{h\phi}|H|^2|\Phi|^2+\f{ \lambda}{2}|\Phi|^4 \,.
\end{align}
The scalar potential contains three dimensionless coupling constants. Later we will see that among them 
$\ld_h$ and $\ld_{h\phi}$ will be almost fixed by two conditions: the weak scale $v\approx 246$ GeV and 
the mass of the SM-like Higgs boson $m_h\approx125$ GeV. Moreover, $\ld$ will be related to $g_{B-L}$ 
via the hidden CW mechanism in the $B-L$ sector. For the sake of EWSB, the coupling constant 
$\ld_{h\phi}$ is assumed to be positive and very small $| \ld_{h\phi} | \ll 1$. 

Let us detail the above picture of $U(1)_{B-L}$ and electroweak symmetry breaking. 
It is well known that the scale invariant version of the SM fails owing to the heaviness of top quark, 
that leads to a maximum rather than a minimum of the CW potential. 
To overcome this problem, one can consider an alternative scenario where CSI spontaneously breaking happens in a hidden 
sector~\cite{hiddenCW}.  This scenario can be realized in a lot of extensions to the SM, e.g., the gauge group extension or dark (hidden) sector extensions.  
In the hidden CW approach, some SM singlet in the hidden sector can first develop nonzero VEV via 
the CW mechanism in the hidden sector, and then transfers this scale to the SM sector through 
the coupling of the singlet scalar to the SM Higgs doublet with a negative coupling constant~\cite{hiddenCW,Iso:2009nw}.
Obviously, the  minimal $U(1)_{B-L}$ model serves as a good example of the hidden CW approach~\footnote{Strictly speaking, $U(1)_{B-L}$ is not hidden because all SM matters are charged under it, but we still adopt this term in the sense of non-obsecration of $Z'$.}.   
We would like to stress that in case of hierarchical VEVs $v\ll v_\phi$,   namely when the  hidden sector 
governs the CSI spontaneously breaking, one can turn off the cross term $\ld_{h\phi}|H|^2|\Phi|^2$  and work in the limit of 
a single scalar field, namely $\Phi$.  This approximation is good  enough in most cases,
except that it may give rise to a discrepancy in the Higgs mixing angle.   
We will come back to this point later.

Let us first study spontaneous $U(1)_{B-L}$ symmetry breaking by the scalar field $\Phi$. 
One can write  $\Phi=(\phi_{\rm cl}+\phi_R+i \,\phi_I)/{\sqrt {2}}$, where $\phi_{\rm cl}$ is the classical 
background, $\phi_I$ is the massless Goldstone boson (GSB) of BLSB and will be eaten by the 
$U(1)_{B-L}$ gauge boson  (denoted as $Z' \equiv Z_{B-L}$ hereafter). While $\phi_R$ is the leftover 
physical  particle  corresponding to the pseudo GSB (PGSB) of CSI spontaneously breaking,  it will play a crucial role in the later  
discussion on dark matter. 

There are different ways to demonstrate the dynamics \'{a} la the CW mechanism, 
and here we follow the original discussion~\cite{CW} which yields the particle mass spectra 
immediately.   In general, the scalar potential at one loop level can be written as
\begin{align}
V(\phi_{\rm cl})=A \phi_{\rm cl}^4+B \phi_{\rm cl}^4\log\f{\phi_{\rm cl}^2}{Q^2},
\end{align}
where $Q$ is the renormalization scale, which is usually taken to be the minimum 
$\langle\phi_{\rm cl}\rangle=v_\phi$ of the scalar potential so as to avoid a large logarithm. 
The logarithmic term  resulting in the scale anomaly is crucial for triggering  CSI spontaneously breaking. 
In fact, the parameter $B$ determines the mass of $\phi_R$:
\begin{align}\label{PGSB:mass}
m_\phi^2=8Bv_\phi^2.
\end{align}
The parameter $B$ receives contributions from any fields that acquire masses by coupling 
to the background field $\phi_{\rm cl}$.  
Specifically to the minimal $U(1)_{B-L}$ model considered in this work, the coefficient $B$ is given by
\begin{align}\label{CW:B}
B=\f{1}{64\pi^2}\L  3\times Q_{\Phi}^4g_{B-L}^4-2\sum_i  (\ld_{N,i}/\sqrt{2})^4  \R,
\end{align}
with $Q_\Phi=2$. This expression will be modified if we introduce a new field  (e.g., the scalar DM 
considered later) that couples to the $\Phi$ field. To ensure $v_\phi$ is a stable minimum point, 
the RHN can not be too heavy:
\begin{align}
B>0\Rightarrow \sum_i\ld_{N,i}^2<\sqrt{6} Q_\Phi^2g_{B-L}^2\Leftrightarrow \sum_iM_{N,i}^2<\sqrt{\f{3}{2}}m_{Z'}^2.
\end{align}
These results are in accord with these of Ref.~\cite{Iso:2009nw} which adopted the viewpoint of 
renomalization group (RG) approach. 

\begin{figure}[htb]
\includegraphics[width=3.5in]{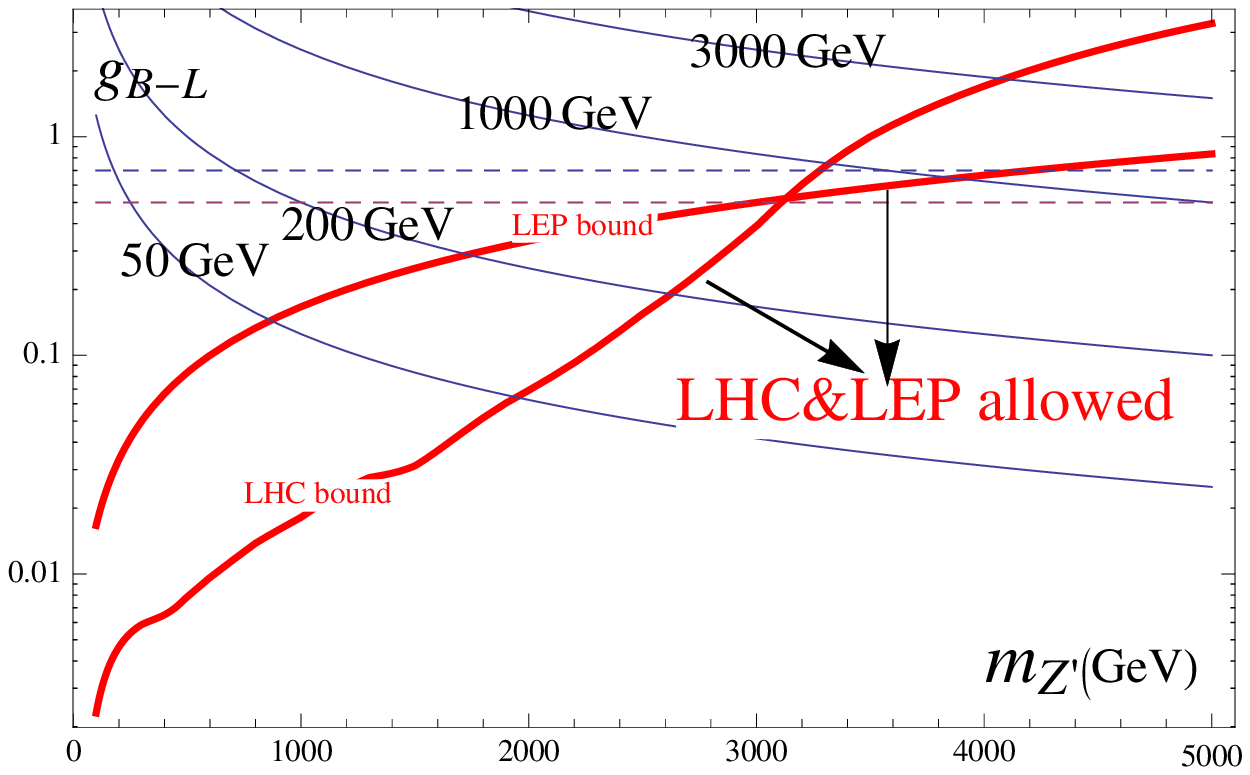}
\includegraphics[width=3.5in]{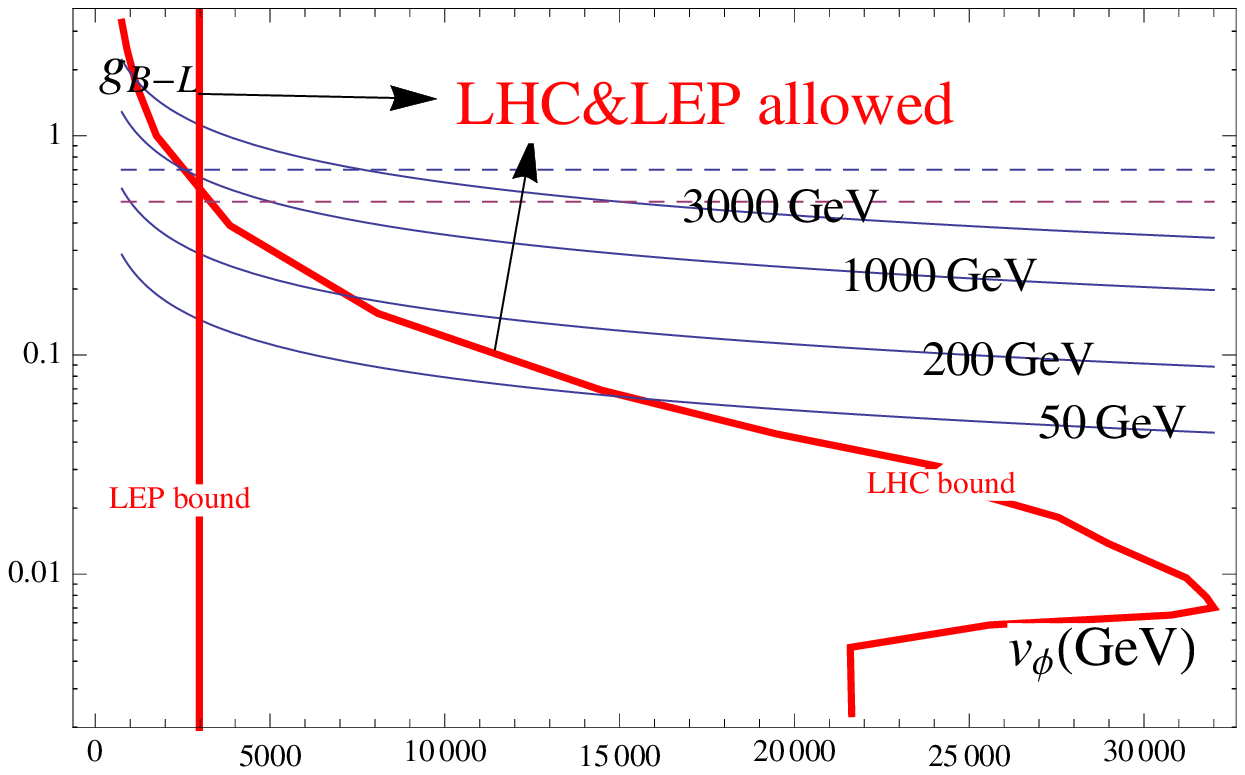}
\caption{Plots of the LHC and LEP constraints on the parameter space of $Z'$ in the MSIBL. 
Upper: exclusion on the $( m_{Z'} , g_{B-L} )$ plane;  
Lower: exclusion on the $( v_\phi , g_{B-L} )$ plane. 
In these plots, the two dashed lines show the upper bounds on $g_{B-L}$, 0.5 and 0.7,  respectively. The curves labeled with 50 GeV, etc., stand for the masses of the PGSB 
$\phi$ without considering the dark matter contribution to CW potential.
}\label{constraint}
\end{figure}

Comments are in order. Firstly, in the above treatment we have turned off the coupling $\ld_{h\phi}$
and reduced the field space into the one-dimensional one. This could be well justified for the case of 
$\ld_{h\phi}\ll1$.  Secondly,  the parameter $A$ is a combination of $\ld$ and  the loop factors: 
\begin{align}\label{CW:A}
A=\f{\ld}{8}+\f{1}{64\pi^2}\left[  3\times Q_{\Phi}^4g_{B-L}^4\L-\f{5}{6}+\ln Q_\Phi^2g^2_{B-L}\R-2\sum_i  (\ld_{N,i}/\sqrt{2})^4\L-\f{3}{2}+\ln \f{\ld_{N,i}^2}{2}\R  \right] .
\end{align}
In dimensional transmutation, the parameter $A$ is related with the parameter $B$ via the minimum 
condition at $v_\phi$, namely $A=-2B$, which allows us to express $\ld$ (at $Q=v_\phi$) in favor of 
$g_{B-L}$ and $\ld_{N}$~\footnote{This expression is a little bit different to that in Ref.~\cite{Iso:2009nw}. 
The basic reason for this difference is that we take a different  renormalization scheme. The results will 
coincide with each other after the redefinition of quartic coupling constant $\ld(v_\phi)\ra \ld(v_\phi)
+Q_\Phi^2\alpha_{B-L}^2\L1+6\ln Q_\Phi^2g^2_{B-L}\R$.   The mass spectrum, which is expressed 
in terms of $g_{B-L}$, is not affected by this  redefinition.}:
\begin{align}\label{ld:v}
\ld(v_\phi)\approx-{6Q_\Phi^4}\alpha_{B-L}^2\L 2-\f{5}{6}+\ln Q_\Phi^2g^2_{B-L}\R.
\end{align}
Thus, $\ld(v_\phi)\sim \alpha_{B-L}^2\ll1$. In case of a very small $\alpha_{B-L}$, which is  
required for a relatively light $Z'$, say around 1TeV, $\ld(v_\phi)$ should be as small as $10^{-10}$. 
This may raise doubt on the single field approximation in the discussion of CSI spontaneously breaking,  because the mixing term 
becomes more important. We will return to it later. Last but not the least,  being the PGSB of CSI spontaneously breaking, $\phi$-mass is suppressed 
by loop factors.   It is illustrative to notice the relation 
\begin{align}
\f{m_{\phi}^2}{m_{Z'}^2}=\f{3 Q_{\Phi}^2\alpha_{B-L}}{2\pi}\L1
- \f{1}{6Q_\Phi^4}\f{\sum_i\alpha_{N,i}^2}{\alpha_{B-L}^2}\R  , 
\end{align}
where we have defined $\alpha_{B-L}\equiv g^2_{B-L}/4\pi$ and $\alpha_{N,i}\equiv \ld_{N,i}^2 /4\pi$. 
Ignoring the RHN contribution \footnote{In the following, we shall assume this simplification in order to 
reduce the number of parameters.}, one can make an estimation of the PGSB mass as follows: 
\[
m_\phi\approx 0.04(g_{B-L}/0.1)\,m_{Z'}  \approx  0.078(g_{B-L}/0.1)^2\,v_\phi . 
\]
It is a few orders of magnitude lighter than $Z'$ and typically lies around the weak scale. 
But in principle this mass can vary within a rather wider region,   from a few GeVs to $\sim$TeVs, 
closely related to the ways to fulfill the LHC bounds on $Z'$. 
In order to see this, we plot several values of $m_\phi$ in Fig.~\ref{constraint}. 

Now we turn our attention to EWSB and consider the change in the Higgs spectrum after 
turning on the cross term. It is a consequence of the second term of Eq.~(\ref{VCW}) that 
produces the negative mass squared term for $H$, i.e., 
$m_{H}^2=\f{\ld_{h\phi}}{2}v_\phi^2>0$  after CSI spontaneously breaking.   It then determines the weak scale $v$ 
to be $v=\sqrt{2m_{H}^2/\ld_h}$ as usual. In turn, we have the relation
\begin{align}\label{mh2}
\ld_{h\phi}v_\phi^2=\ld_h v^2\approx m_h^2,
\end{align}
with $m_h$ the SM-like Higgs boson mass, obtained in the approximation of neglecting the 
mixing effect. This relation helps us to reduce one parameter of the model, $\ld_{h\phi}$, 
for  convenience. If we adopt the Gildner-Weinberg approach~\cite{GW} that is suitable for 
handling the CW mechanism in the multi-field space, we would get the same relation 
via determining the so-called flat direction $\vec \varphi= \varphi \vec n$ with 
$\vec n=(n_\phi,n_h)$ at tree level:
\begin{align}
\f{h_{\rm cl}^2}{\phi_{\rm cl}^2}=\f{n_h^2}{n_\phi^2}=\f{\ld_{h\phi}}{\ld_h}=\f{\ld}{\ld_{h\phi}}\ll1,
\end{align}
where  $\phi_{\rm cl}=\varphi n_\phi$ and $h_{\rm cl}=\varphi n_h$ with $\varphi$ determined at loop level. Note that the components of the PGSB ${\cal P}$, just like $\vec n$, are also determined at tree level.  In practice, they are 
nothing but the fractions $n_h$ and $n_\phi$, i.e., ${\cal P}=n_h h+n_\phi \phi_R$. For later 
convenience, let us define two mixing angles $\theta$ and $\beta$ as
\begin{align}\label{PGSB:mix}
\tan\theta=\f{n_h}{n_\phi}=\f{v}{v_\phi} = \frac{1}{\tan\beta} \,.
\end{align}
%We also define $\tan\beta\equiv v_\phi/v=\cot\theta$. 
Then the mixing angle $\theta$ becomes very small, $\theta \lesssim{\cal O}(0.1)$ for $v_\phi>3.5$ TeV. 

A few comments are in order. Firstly, let us notice that the above expression is different from  
the one derived in Ref.~\cite{Iso:2009nw} where $\tan\theta=\cot\beta/(1-m_\phi^2/m_h^2)$. 
%As one can see, 
Note that they coincide with each other only in the limit $m_\phi^2\ll m_h^2$ with 
$m_\phi^2$ given by Eq.~(\ref{PGSB:mass}). In particular, the mixing angle $\theta$ will blow up 
in their expressions as two masses become degenerate. Since the PGSB components should 
be related to the ratio of two VEVs only, Eq.~(\ref{PGSB:mix}) is a more precise expression. Secondly, in this approach, the aforementioned extremely small $\ld$ problem does not bother us because we are now considering a complete potential. Finally, we mention that in this model the vacuum stability problem in the SM is almost not 
affected, because the off-diagonal entry of the mass matrix is tiny,  and hence the Higgs mixing effect merely gives a tiny shift to the diagonal elements. In other words, Eq.~(\ref{mh2}) works quite well to determine $\ld_h$. In what follows we will denote ${\cal P}$ as $\phi$.

%---------------------------------------------------------------------------------------------------------
\subsection{$Z'$ and $\phi$ at colliders}
%---------------------------------------------------------------------------------------------------------

The minimal $U(1)_{B-L}$ model predicts two new particles $Z'$ and ${\phi}$ that can be searched  
for at current/future colliders. Actually, the current LHC bound on $Z'$ is quite stringent, 
because both quarks and leptons are charged under the $U(1)_{B-L}$ gauge group. 
It has immediate implications for DM phenomenology too. 
Thus in this subsection we will give a up-to-date analysis on the constraints on $Z'$. 
We will also briefly comment on the prospect on the search for ${\phi}$.

Firstly, LEP II  measured the cross section for $e^+e^-\ra \bar ff$ ($f\neq e$) above the $Z$-pole, 
which yielded the following constraint on the heavy $Z'$~\cite{Carena:2004xs,Heeck:2014zfa}:  
\begin{align}
m_{Z'}/g_{B-L}=Q_{\Phi}v_\phi\geq 6.9{\rm\,TeV} \,.
\end{align}
Then one can immediately get a lower bound $v_\phi\geq 3.5{\rm\,TeV}$. Next, the current 
CMS and ATLAS searches for dilepton resonances give a bound on the $Z^{'}$ through the Drell-Yan  
processes ($q\bar q\ra Z'\ra \bar \ell \ell$, with $\ell=e$ or $\mu$)~\cite{CMS}. 
The bound is derived in the $( m_{Z'} , \sigma \cdot {\rm Br}) $ plane, with $\sigma$ and $\rm Br$ 
denoting for  $\sigma(pp\ra Z')$ and ${\rm Br}(Z'\ra \bar \ell \ell)$, respectively. 
We can calculate the $Z^{'}$ production  cross section at $pp$ collision and the branching ratios for 
$Z^{'}$ decays into the SM fermions in the MSIBL, and then compare their product to the experimental 
upper bound~\footnote{Note that the $Z'$ involved in LEP II search is off-shell and thus the corresponding 
bound depends only on cross section. This is different from the LHC case, which involves the on-shell 
$Z'$ and thus is additionally sensitive to the branching ratio. 
Thus the LHC bound can be relaxed by opening new decay channel for $Z'$.}. 
In this way, we are able to obtain an exclusion plot in the $m_{Z'}-g_{B-L}$ plane, which is of interest in 
DM phenomenology.  From Fig.~\ref{constraint},  we have two ways to avoid the LHC and LEPII constraints: 
One way is to assume a lighter $Z'$ and smaller $g_{B-L}$; the other way is to assume a rather heavy $Z'$ 
which allows a larger $g_{B-L}$. Obviously,  in the former case the LHC bound turns out to be more stringent 
than the LEPII bound, while in the latter case the LEPII bound becomes more powerful.  
Note that imposing the upper bound on $g_{B-L}\leq0.5$ yields a similar bound to the LEPII bound 
(in the heavy $Z'$ region).  We also show the LHC and LEPII exclusion plots in the $( v_\phi , g_{B-L})$ 
plane in Fig.~\ref{constraint}.

\begin{figure}[htb]
\includegraphics[width=3.0in]{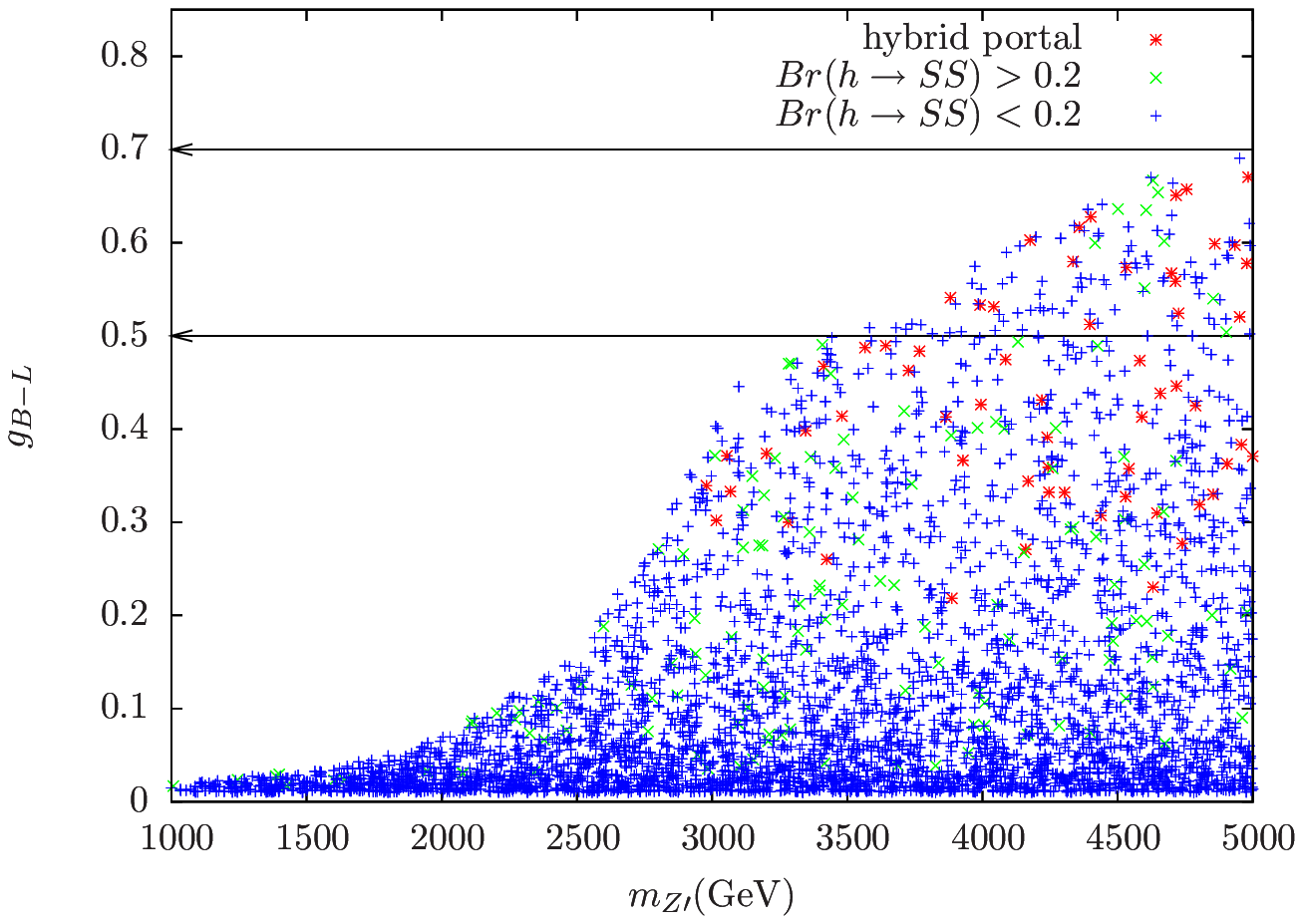}
\includegraphics[width=3.0in]{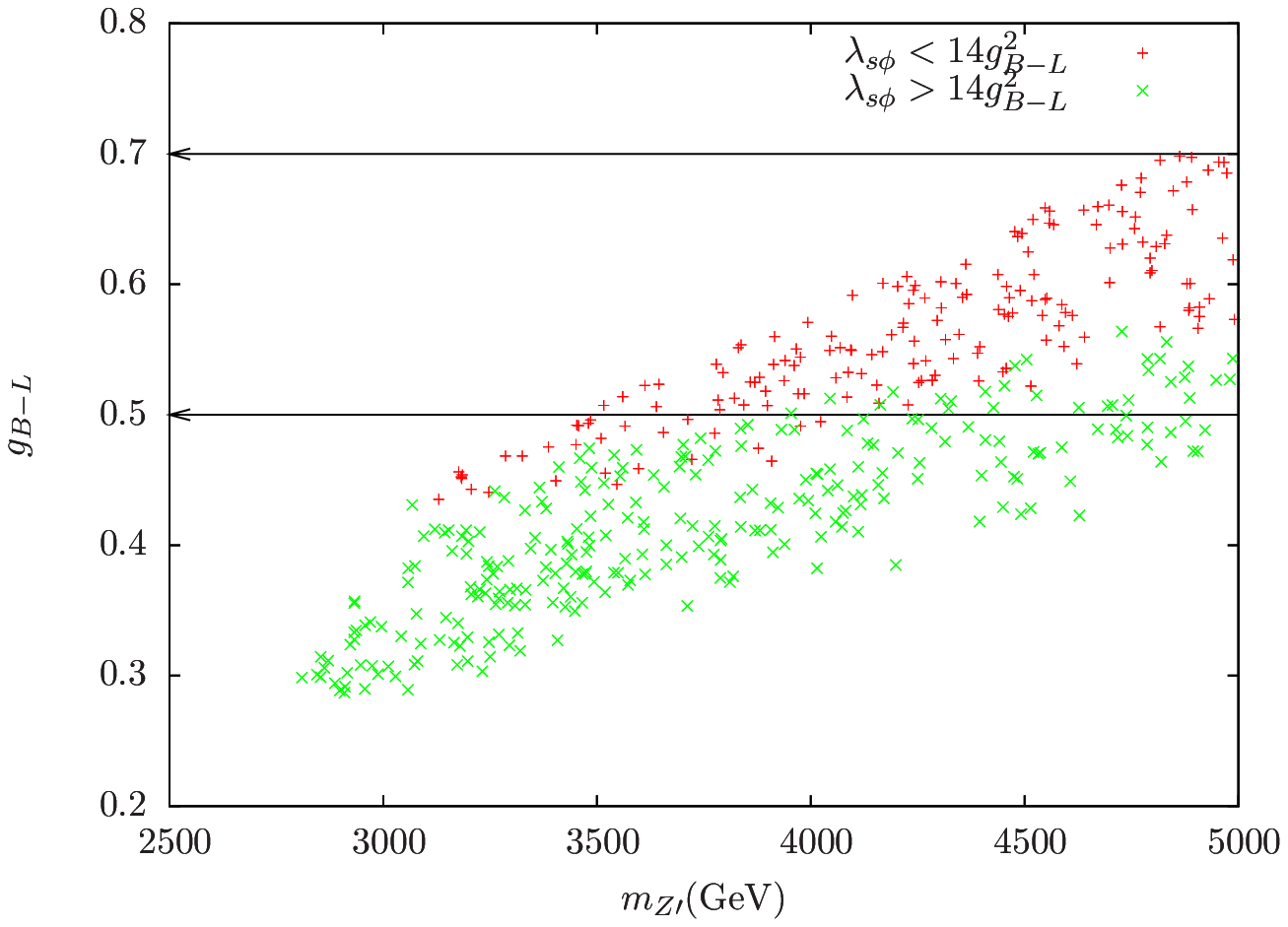}
\includegraphics[width=3.0in]{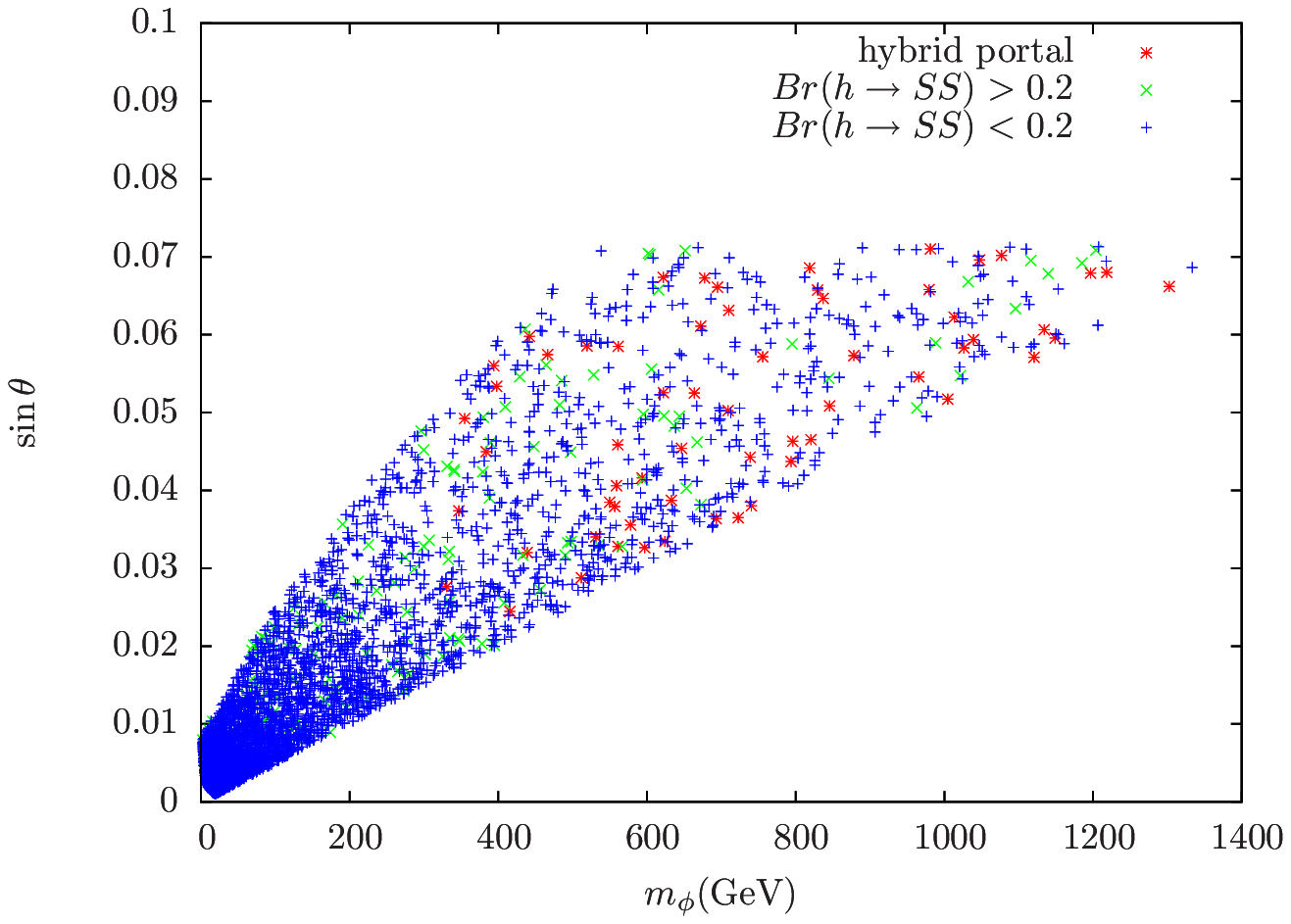}
\includegraphics[width=3.0in]{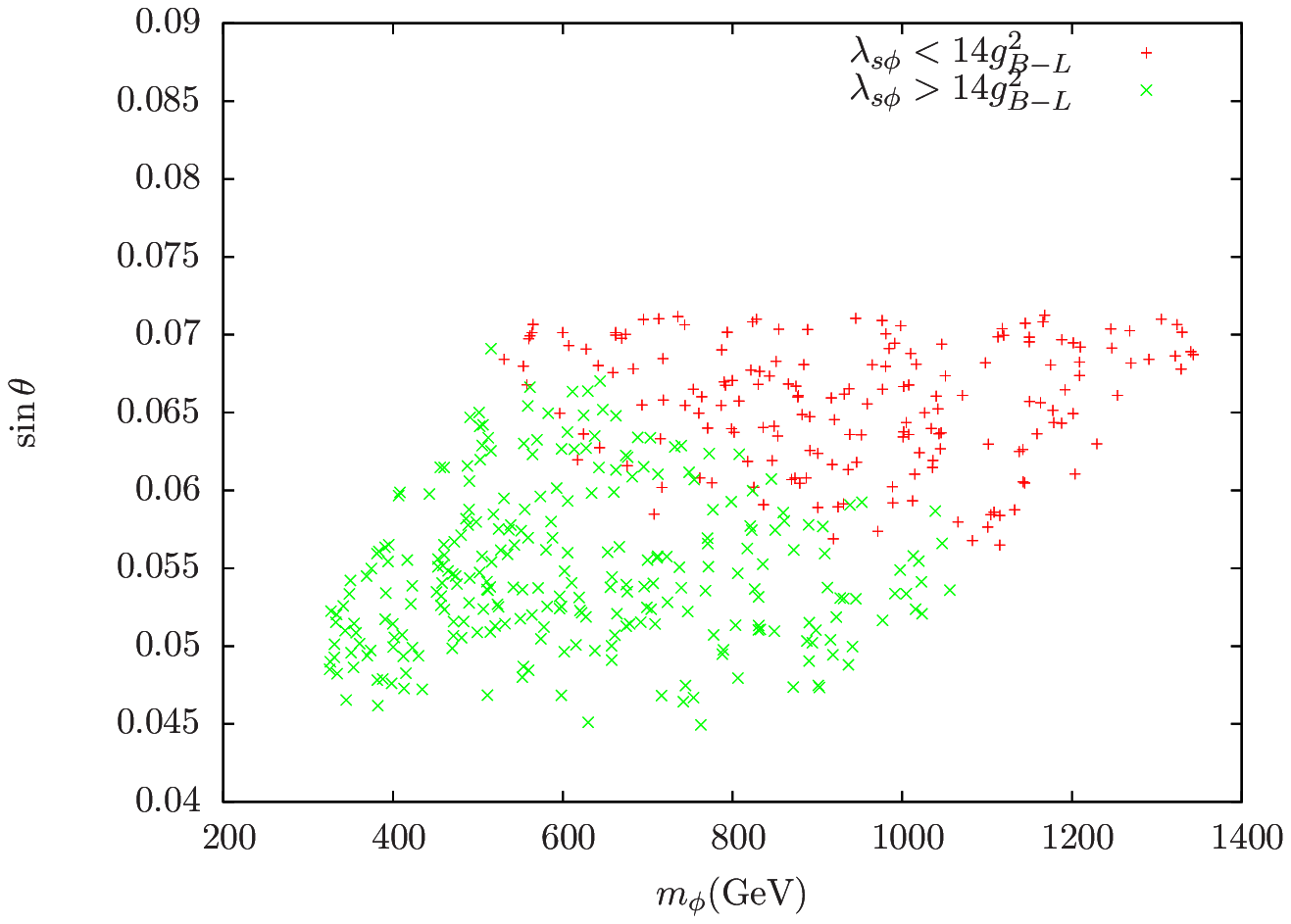}
\caption{Surviving spectrum. All the points satisfy $0.09<\Omega h^2<0.12$ and are allowed by LUX. Top (bottom) left: $Z'$ ($\phi$) in the Higgs portal scenario; Top (bottom) right: $Z'$ ($\phi$)  in the dark Higgs portal scenario. The perturbativity bounds on $g_{B-L}$ is schematically presented in two cases, i.e., the line $g_{B-L}=0.5$ and 0.7.}\label{spectrum}
\end{figure}

The PGSB ${\phi}$ is another prediction of the MSIBL. But hunting for this new particle 
at colliders is not promising because the mixing angle is very small 
$\sin\theta\lesssim{\cal O}(0.1)$. We display distributions of ${\phi}$ in the 
$( m_\phi , \sin\theta )$  plane in Fig.~\ref{spectrum}. $m_\phi$ can vary in a wide region 
(in particular in the Higgs portal scenario), from GeV to TeV. In the relatively heavier region 
where $\sin\theta$ is comparatively larger, the most feasible method is to observe a pair of boosted 
$W$'s from the resonant production of ${\phi}$: $GG\ra {\phi}\ra W^+W^-$. 
For illustration, at the  LHC@14 TeV and for $m_\phi=1$ TeV and $\sin\theta=0.07$, 
the production cross section of the boosted $W^+W^-$ is $\sim{\cal O}(fb)$. 
In the highly decoupling region with very small $\theta$, 
${\phi}$ can be produced associated with $Z'$ via 
$\bar qq\ra {Z'^*}\ra Z' {\phi} $. 
However, considering the strigent bound on $Z'$,  the production rate should be tiny.

%---------------------------------------------------------------------------------------------------------
\section{Classification of Accidental dark matter in MSIBL}
%---------------------------------------------------------------------------------------------------------

In this section we consider embedding the idea of accidental DM (aDM) in MSIBL model. 
With the aid of CSI and local $U(1)_{B-L}$  symmetry,  DM can be realized without much difficulty. 
Based on the MSIBL,  candidates for aDM can be classified into the following four cases: 
\begin{itemize}
%\begin{enumerate}
\item Case A: aDM is a singlet with respect to all gauge groups, and could be a singlet  scalar 
$S$ or a singlet fermion $\chi$.  However, $\chi$ must be massless in the MSIBL (which offers no singlet scalar with VEV to give mass for $\chi$), so we do 
not consider this case furthermore. Then $S$ becomes the unique candidate. It conserves 
an accidental $Z_2$ (see Sec. IIIA for detailed discussions). 
\item Case B:  aDM is a scalar $S_X$ carrying $B-L$ charge $X$. The trivial case with 
a peculiar $X$ such that it couples to other fields only via $|S_X|^2$ will be of no interest. 
Then the unique option is $X=\pm2/3$, giving rise to accidental $Z_3$ DM 
(see Sec. IIIB for detailed discussions).  Fermionic candidates suffer from the gauge anomaly,  
which probably ask for a stack of new particles, a case with much less attraction (the same 
arguments apply to the third and fourth cases below). 
\item Case C: aDM is charged under $SU(2)_L\times U(1)_Y$, but electrically neutral. 
It is the neutral component of a multiplet $(2j+1,Q_Y)$  with $j$ an integer or half integer. 
As shown in Ref.~\cite{Guo:2014bha}, the lower $j\leq 1$ cases fail in providing aDM. For example, 
a triplet scalar $T=(3,0)$ allows a term $\bar L TL$ with $L$ the lepton doublet. Although the higher $j\geq 2$ case is viable, it is the SM gauge symmetries instead of SI and 
local  $U(1)_{B-L}$ that guarantee the accidental $Z_2$. Such models actually become 
the minimal DM  scenario proposed by Strumia et al.~\cite{Cirelli:2005uq}.  
\item Case D: aDM comes from a multiplet with double charges, i.e., $(2j+1,Q_Y,Q_{B-L})$. 
For example, the $j=1/2$ case may lead to an inert Higgs doublet.  
However, unlike the Case (B), here we do not have a way to fix the quantum number $Q_{B-L}$. 
\end{itemize}
%\end{enumerate}
In summary, (A) and (B) give the simplest and the most relevant aDM candidates within MSIBL models. 
In what follows we will study them one by one in two separating subsections. 
    
%---------------------------------------------------------------------------------------------------------
\section{Real singlet aDM with accidental $Z_2$}\label{Real:Z2}
%---------------------------------------------------------------------------------------------------------

A real singlet scalar $S$ can be accidentally stable in the MSIBL model, which is explicitly seen in its most general interacting Lagrangian \footnote{The model Lagrangian  becomes automatically renormalizable due to CSI, if we use the dimensional regularization with minimal subtraction.}
\begin{align}\label{sphi}
{\cal L}_{DM}= - \frac{1}{2}\lambda_{sh}S^2(H^{\dag}H) - 
\frac{1}{2}\lambda_{s\phi}S^2(\Phi^{\dag}\Phi) - \f{\ld_s}{4}S^4,
\end{align}
where $S$ is odd under a $Z_2$ symmetry, which is automatic or accidental as a result of the 
field content and symmetries furnished by the model.  It is important to note that
this feature is not broken by quantum corrections.  Quantum corrections generate operators which are
even powers of $S$ and $Z_2$ is not broken at loop level. 
It is crucial that RHNs carry $U(1)_{B-L}$ charge so that the terms $SN_i^2$ terms  violating the accidental 
$Z_2$ symmetry are forbidden
~\footnote{In Ref.~\cite{Guo:2014bha} where $U(1)_{B-L}$ is not gauged, and hence an exact 
accidental $Z_2$ is problematic after incorporating RHN to interpret neutrino masses. 
Other advantages having gauged $U(1)_{B-L}$ are that  it does not incur large coupling around 
weak scale, and successful DM phenomenology  can be accommodated much more readily 
in the presence of dark Higgs portal $S^2|\Phi|^2$.}. The self coupling of $S$ ($\ld_s$) is irrelevant to DM phenomenologies but relevant to guarantee that the vacuum under consideration is a global minimum, or equivalently dark matter parity $Z_2$ is not spontaneously breaking. For that sake, $\ld_s$ should not be very small (says much smaller than 0.1) in the light of the discussion in Appendix~\ref{global}. However, viewing from RGE flow, it is in strong tension with a large cross term of the two scalars in ${\cal L}_{DM}$. Eq.~(\ref{lsRGE}) implies that if $\ld_{s\phi}\gg 1$, it will decrease $\ld_s$ fast and make it negative at some scale not far from the boundary $\Ld$.

\begin{figure}[htb]
\includegraphics[width=3.23in]{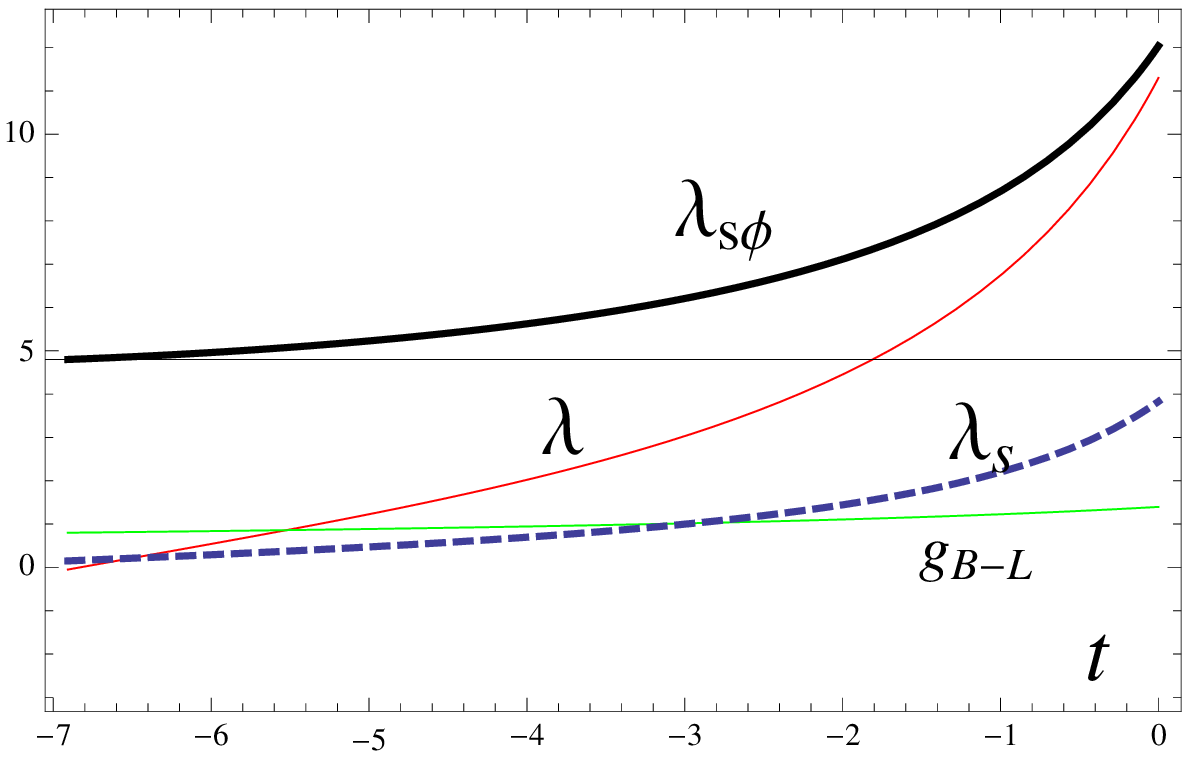}
\includegraphics[width=3.2in]{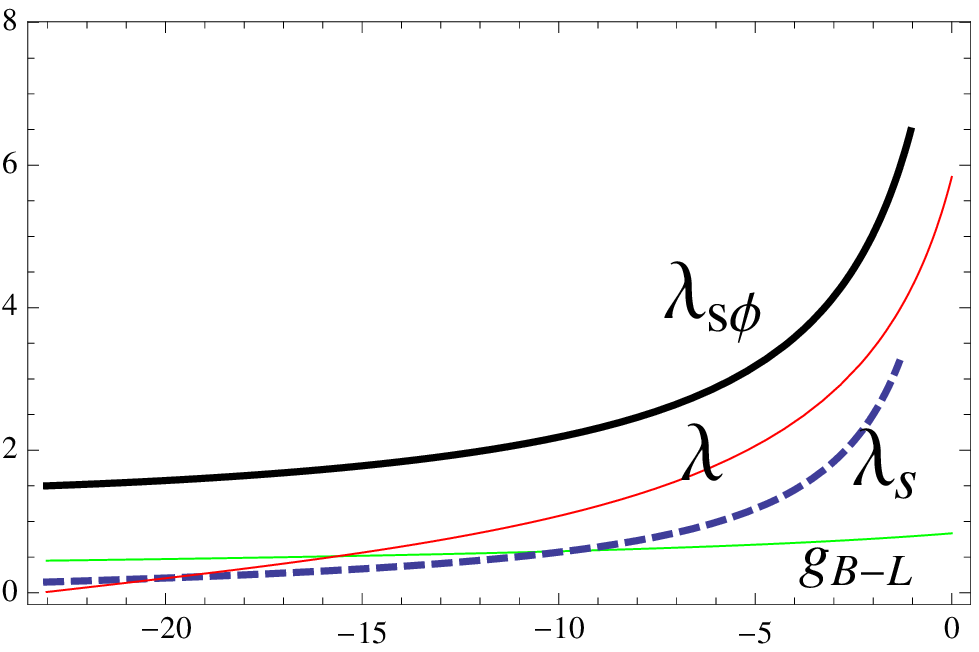}
\caption{RGE flows of four couplings with a large $\ld_{s\phi}(v_\phi)$, left: $\Ld=10^3v_\phi$; right: $\Ld=10^{10}v_\phi$. The perturbativity upper bound at $\Ld$ for the quartic couplings and $g_{B-L}$ are $4\pi$ and $\sqrt{4\pi}$, respectively.}
\label{pic:RGE}
\end{figure}

The large cross term endangers vacuum stability from another aspect. From the structure of $\beta_{\ld_{s\phi}}$ in Eq.~(\ref{ru:sphi}), it is clearly seen that, to allow $\ld_{s\phi}(v_\phi)$ as large as 5.0 after running from a high scale $\Ld\gg \rm TeV$, one has to rely on a large $g_{B-L}(\Ld)$ that can substantially slow the reduction of $\ld_{s\phi}(t)$ from $t_0=0$ to $t_\Ld=\log (v_\phi/\Ld)$. Then $\ld$, the quartic coupling of $\Phi$, is going to be quickly driven negatively large through the large $g_{B-L}^4-$term in $\beta_{\ld}$, see Eq.~(\ref{ru:ld}). This is not consistent with the dimensional transmutation condition Eq.~(\ref{ld:v}), which means $|\ld(v_\phi)|\ll1$. In summary, to make the vacuum with only $\Phi$ acquiring VEV absolutely stable, we require either fairly weak couplings or a sufficiently low scale $\Ld$~\footnote{The most convincible $\Ld$ is $M_{\rm Pl}$. Then we may say that, in the absence of intermediate thresholds, the model is free of hierarchy problem even with gravity. Quantum gravity may leave SI in the IR.}. In Fig.~\ref{pic:RGE} we show two numerical samples of RGE flow with a low (left) and high (right) scale $\Ld$, respectively. In each case, we have tried to get a maximal $\ld_{s\phi}(v_\phi)$ by adjusting $g_{B-L}$ and $\ld_s$. At $v_\phi$, it is required that $\ld_s\gtrsim0.1$ and $|\ld|\ll1$.

Asides from stability, DM models with CSI shed light on understanding another basic property 
of DM, namely the origin of its mass~\cite{strong,Guo:2014bha,common}.   When specified to the MSIBL, 
aDM mass scale originates from CSI spontaneously breaking in the $U(1)_{B-L}$ sector, just like the EW scale. 
More exactly, aDM actually acquires mass from two dynamical sources: 
\begin{align}\label{DM:mass}
m_{S}^2=\frac{1}{2}(\lambda_{sh}v^2+\lambda_{s\phi}v_\phi^2) .
\end{align}
However, in most of the interesting parameter space, practically the second source always dominates  
because of the hierarchy $v\ll v_\phi$ and moreover $\ld_{sh}\ll1$ from the DM direct detection bounds. 
In principle, DM mass can vary in a vast region, from a few GeVs to TeVs  depending on the quartic 
coupling $\lambda_{s\phi}$ (and $\lambda_{sh}$ in case of very small $\lambda_{s\phi}$).   

Concerning thermal relic density of dark matter, there are two main distinct scenarios for this 
real singlet DM. One is the well known Higgs portal in which $\ld_{sh}-$term plays the crucial 
role while the $\ld_{s\phi}-$term merely provides a (major part) mass of DM. One can give the crude condition for the need for Higgs portal. It is obtained simply by requiring the annihilation 
rate of $SS\ra {\phi} {\phi}$ is short for the typical value, namely $ \sigma_{{\phi} {\phi}}v\lesssim$ 1 pb. The analytical expression will be given later in Eq.~(\ref{SS2phi}), from which  
one can infer that one has to fall back on the Higgs portal when 
$\ld_{s\phi}\lesssim 1.0\times \L \f{5\rm\,TeV}{v_\phi}\R^2$. 
By contrast, the other possibility is realized  in the vanishing $\ld_{sh}$ limit, for which the 
single $\ld_{s\phi}-$term accounts for both DM mass generation and DM annihilation (into 
the PGSB pair) simultaneously. This is a well expected scenario since the PGSB can be 
naturally as light as the weak scale. We dubbed this as the dark Higgs portal 
scenario. But practically this scenario is not dark and it can be directly detected sooner or  later. 
Both of these scenarios are simplified, if we decouple the two Higgs sectors, but take into 
account their small mixing, which is compensated by the large $v_\phi$, makes a difference. It may open new possibilities and we will comment on them. In what follows we will go into the details of each scenario respectively. 

%---------------------------------------------------------------------------------------------------------
\subsection{Higgs portal: an old story}
%---------------------------------------------------------------------------------------------------------

Although Higgs portal DM has been studied extensively, our arguments to arrive this kind of 
DM model  shows more theoretical attraction, as stated before. Here we only brief its status 
facing the bounds from the DM direct detection and Higgs invisible decay, which is stringent 
once kinematically allowed. As for its confronting with the indirect detections or global survey, we refer to~\cite{Higgsportal}.   
In this case, the SM Higgs boson $h_{\rm SM}$ mediates a large DM-proton spin-independent 
(SI) scattering which is given by % with a cross section
\begin{align}\label{SI:H}
\sigma^{(n)}_{\rm SI}=\frac{\lambda_{sh}^2}{4\pi}\f{ \mu_n^2 m_n^2}{m_S^2m_h^4}\L\sum_{q=u,d,s} f^{(n)}_{T_q}+\frac{2}{9}f^{(n)}_{T_G}\R^2,
\end{align}
where $\mu_n=m_n m_S/(m_n+m_S)\approx m_n$ is the reduced mass of DM and proton. 
The values of the nucleon parameters $f^{(n)}_{T_q}$ etc., 
can be found in Ref.~\cite{Gao:2011ka}.    
In this paper we take the value in the bracket, defined as $f_N^{(n)}$ hereafter, to be 0.3 for estimation. 

\begin{figure}[htb]
\includegraphics[width=4.1in]{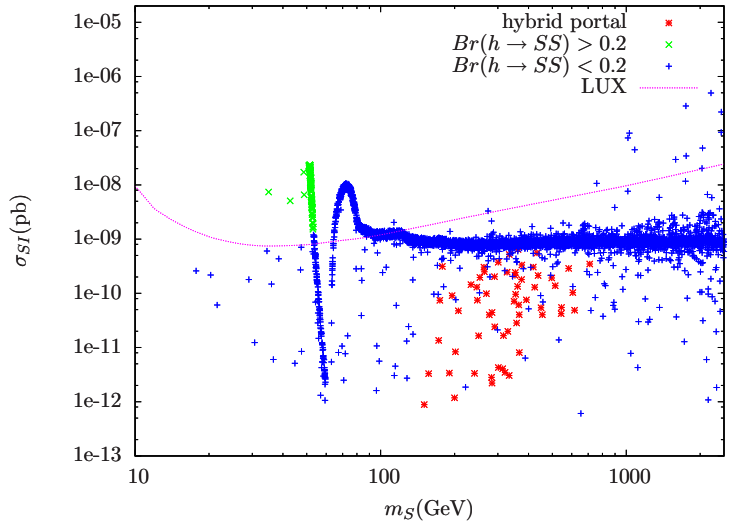}
\caption{Status of dark matter under the currently most stringent bound from DM direct search, LUX~\cite{LUX}. In this plot all the points have good relic density, namely have $0.09<\Omega h^2<0.12$. The points in the Higgs portal scenario are around the blue band.  While the red points are samples from the mixing effect with $m_\phi$ near $2m_S$, see details in the text.
}
\label{Higgsp}
\end{figure}

DM direct detection experiments like XENON100 and especially the recent LUX 
~\cite{LUX}   place an upper bound on the combination $\ld_{sh}/m_S$. 
For instance, for a TeV scale heavy DM one has $\sigma_{\rm SI}\lesssim 10^{-8}\,\rm pb$, 
and hence the upper bound on $m_{S}/\ld_{sh}\gtrsim 1.0$ TeV. On the other hand, this factor 
enters the annihilation cross section of the TeV scale DM, i.e., $\sigma v \simeq (\ld_{sh}/m_S)^2/32\pi\lesssim 10^{-8}\rm GeV^{-2}\sim{\cal O}(1)$ pb,  which is just at the correct 
order of magnitude.  Thus the heavier DM mass region is still allowed, given a relatively large 
$\ld_{sh}$. But this region is expected to be closed in the coming years. In the lighter DM 
region below $m_h$, by virtue of the Higgs resonance, there is a narrow room around 
$m_h/2$ that is hard to be closed by direct detections. Higgs invisible decay yields a loose constraint, if we naively require its branching ratio is less than $20\%$~\cite{Belanger:2013kya}. To completely rule out this region one has to rely on indirect detections. We show the results in Fig.~\ref{Higgsp}, which is obtained with the help of numerical package Micromega 3.2~\cite{Belanger:2013oya}, scanning the parameter space shown in 
Table~\ref{para1}. 
\begin{table}[h]
 \small
 \begin{center}
  \begin{tabular}{|c|c|c|c|c|c} \hline
  $m_{Z^{'}}$ & $\alpha_{B-L}$ & $\lambda_{sh}$ & $\lambda_{s\phi}$   \\ \hline
   [0.5,\,5] TeV & [$1\times 10^{-4}$,$5\times 10^{-3}$] & [0,1.0] &[$1\times 10^{-5}$,\,$6.0$] \\ \hline
  \end{tabular}
  \caption{\label{} Scanned parameters. }\label{para1}
 \end{center}
\end{table}

%---------------------------------------------------------------------------------------------------------
\subsection{Dark Higgs portal: DM as a trigger of CSI spontaneously breaking}
%---------------------------------------------------------------------------------------------------------

Now we turn our attention to the dark Higgs portal scenario. 
For simplicity, we turn off $\lambda_{sh}$ for qualitative discussion. 
In the actual numerical analysis we will turn it on and scan over it. 
As stressed before, in most case $SS\ra {\phi} {\phi}$ dominates over other annihilation modes. 
The main contributions to this annihilation come from three Feynman diagrams: 
The contact vertex, the $S-$channel mediated by ${\phi}$ and the $t/u$-channel mediated 
DM itself. 
The amplitude is given by
\begin{align}
{\cal M}_{SS\ra {\phi} {\phi}}\simeq - 2\frac{\lambda_{s\phi}^2v_\phi^2}{m_{\phi}^2-2m_S^2} -  \lambda_{s\phi}.
\end{align}
We have worked in the non-relativistic limit so that the $t-$ and $u-$channel contributions are equal and independent on the scattering angle. The annihilation cross section times the relative velocity is given by 
\begin{align}\label{2phi}
 \sigma_{{\phi} {\phi}}v\simeq\f{1}{2}\times\frac{1}{8\pi}\frac{1}{4m_S^2}|{\cal M}_{{\phi} {\phi}}|^2{\sqrt{1-r_{\phi}}}\simeq \f{1}{32\pi} \f{\ld_{s\phi}}{v_\phi^2}\L\f{2+r_\phi}{2-r_\phi}\R^2\sqrt{1-r_{\phi}}.
\end{align}
with $r_\phi\equiv m_{\phi}^2/m_S^2$. In the limit $r_\phi\ll1$ which works well in this 
scenario, the final expression is simplified to 
\begin{align}\label{SS2phi}
\sigma_{{\phi} {\phi}}v\approx \f{1}{32\pi }\f{\ld_{s\phi}}{v_\phi^2}=\f{1}{16\pi }\f{m_S^2}{ v_\phi^4} .
\end{align}
The second equation was obtained by using an approximation,  
$m_S^2\approx \ld_{s\phi}v_\phi^2/2$. 
Owing to the common origin for DM mass and annihilating dynamics, the cross section is 
quite sensitive to VEV $v_\phi$. A relatively low scale $v_\phi$ and moreover a larger $\ld_{s\phi}$ are needed. For instance, even if $v_\phi$ is around  $\gtrsim$ 3.5 TeV derived from the LEP II experiments, 
we still need $\ld_{s\phi}\approx 3.0$ in order to produce the correct thermal cross section 
$\sigma_{{\phi} {\phi}}v\simeq $ 1pb. Recall that low $v_\phi$ is allowed only 
in the heavier $Z'$ region. Thus the viable dark Higgs portal scenario is supposed to dwell 
in such a slice of parameter space: $v_\phi\gtrsim 3.5$ TeV, $g_{B-L}\gtrsim 0.1$ and  
$\ld_{s\phi}\gtrsim 1.0$, which is shown in the upper panel of Fig.~\ref{dHiggs}. It is clear that the LEPII and 
perturbativity bounds could help to rule out a large portion of the parameter space, and in the plot we merely show points satisfying $v_\phi\geq3.5$ TeV and $g_{B-L}\leq0.7$.

We would like to pause here to discuss perturbativity bounds on two crucial parameters $g_{B-L}$ and $\ld_{s\phi}$. At one-loop, the evolution of gauge coupling $g_{B-L}$ is independent on other couplings and can be solved analytically, 
\begin{align}\label{}
g_{B-L}(t)=\f{1}{\L- 3t/2\pi^2+4g^{-2}_{B-L}(\Ld)\R^{1/2}}. 
\end{align}
Thus, given a sufficiently long RGE course $g_{B-L}(0)\lesssim 2.5/\sqrt{-t_\Ld}$. For instance, for $\Ld\gtrsim10^{7}v_\phi$ we have $g_{B-L}(v_\phi)\lesssim0.6$. As for $\ld_{s\phi}$, we have shown previously its largeness may render vacuum unstable. But if $\Ld$ can be fairly low, says 1000 TeV, $\ld_{s\phi}(v_\phi)$ around 5.0 is still allowed. In this paper we are not restricted to $\Ld=M_{\rm Pl}$ and accept a relatively low $\Ld$, which justifies the discussions of this subsection.

\begin{figure}[htb]
\includegraphics[width=3.8in]{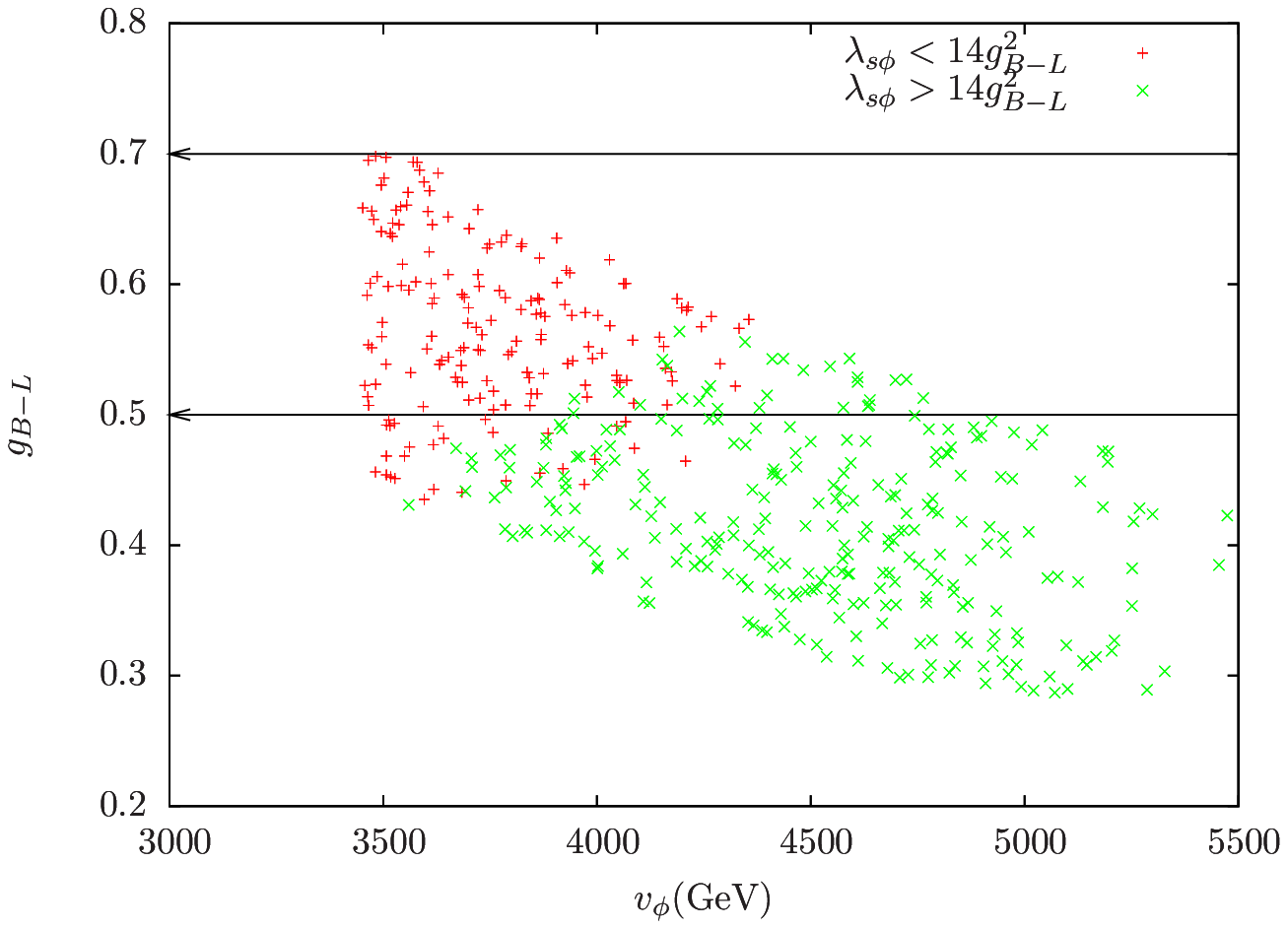}
\includegraphics[width=3.9in]{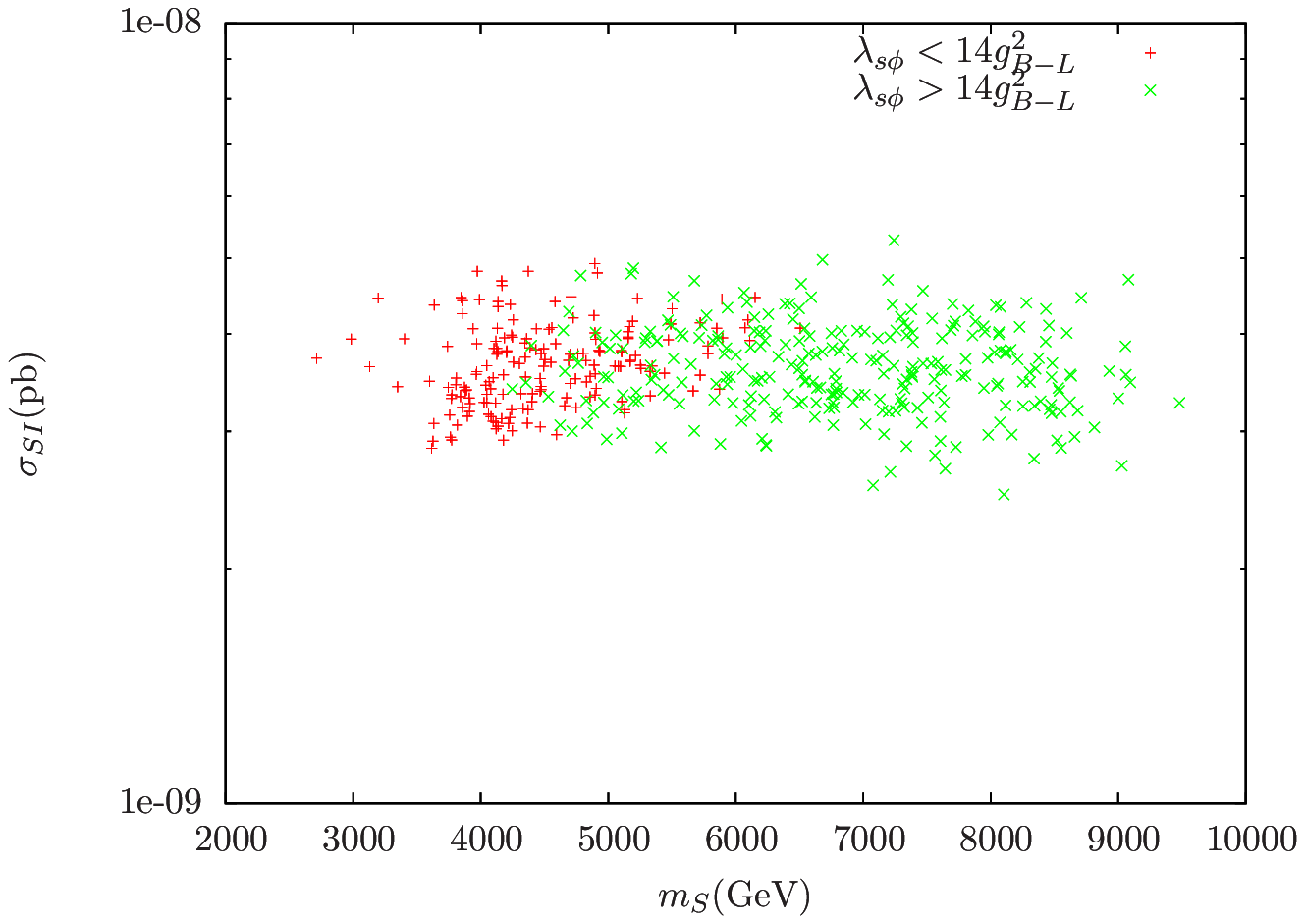}
\caption{Dark Higgs portal scenario on the $v_\phi-g_{B-L}$ plane. The LEP II excludes the region with $v_\phi<3$ TeV. The green and red points belong to the scenarios where DM and $Z'$ dominantly trigger CSI spontaneously breaking, respectively. We have imposed a perturbativity upper bound $\ld_{s\phi}\leq6.0$ in numerical scanning.}\label{dHiggs}
\end{figure}

%By contrast, in the case of a negligible $g_{B-L}$, at low energy we can only have  $\ld_{s\phi}$ smaller than 1, which is too small to account for the thermal cross section Eq.~(\ref{SS2phi}), in particular for a larger $v_\phi$. For example, if $v_\phi=5$ TeV, $\ld_{s\phi}=6.4$ is needed. In summary, from Fig.~\ref{dHiggs} it is seen $g_{B-L}\gtrsim 0.3$, and thus $g_{B-L}$ should be large at $\Ld$. This means that the $\ld_{s\phi}(\rm TeV)\leq 6.0$ used before is reasonable. 

\begin{figure}[htb]
\includegraphics[width=3.8in]{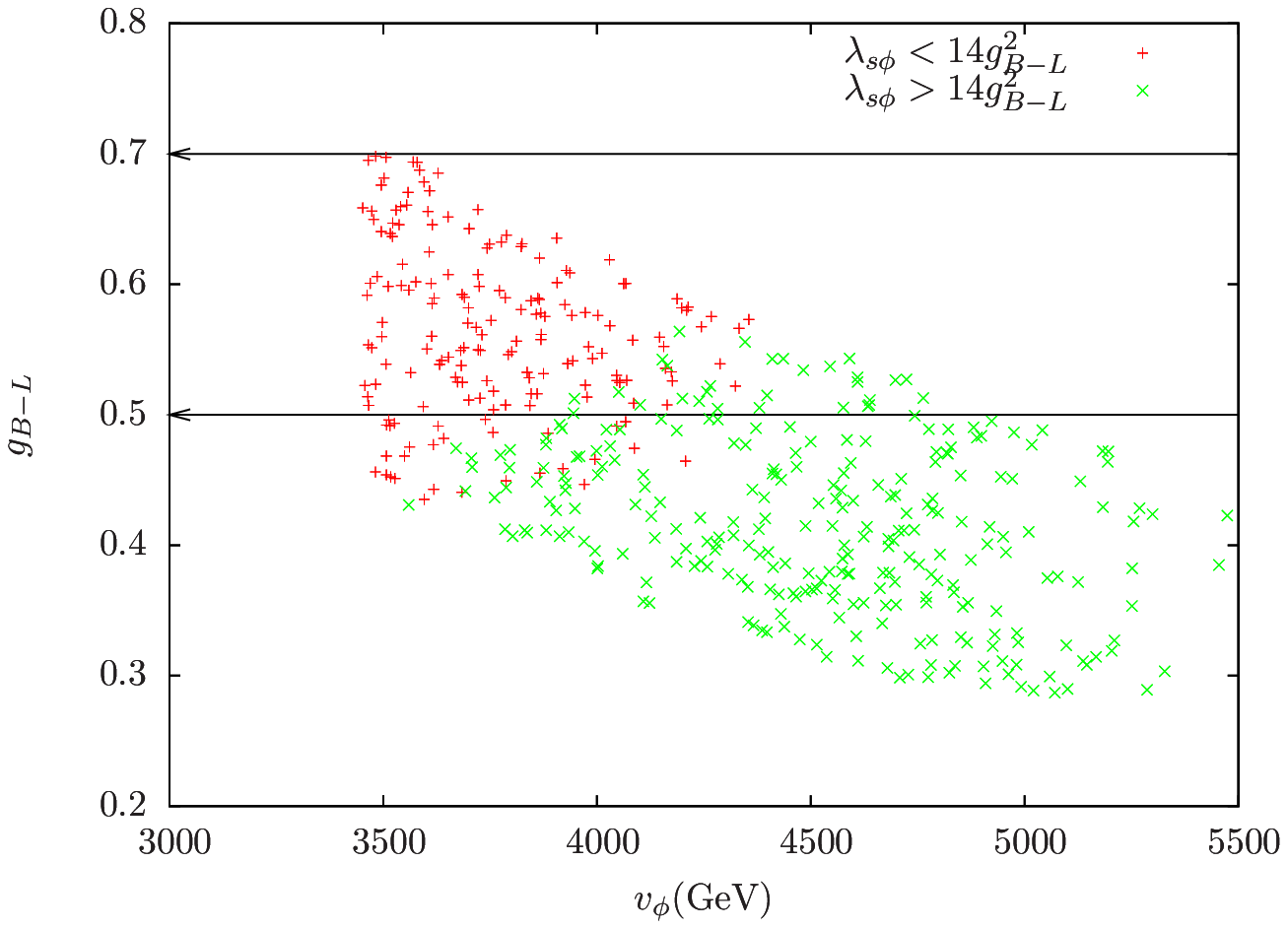}
\includegraphics[width=3.9in]{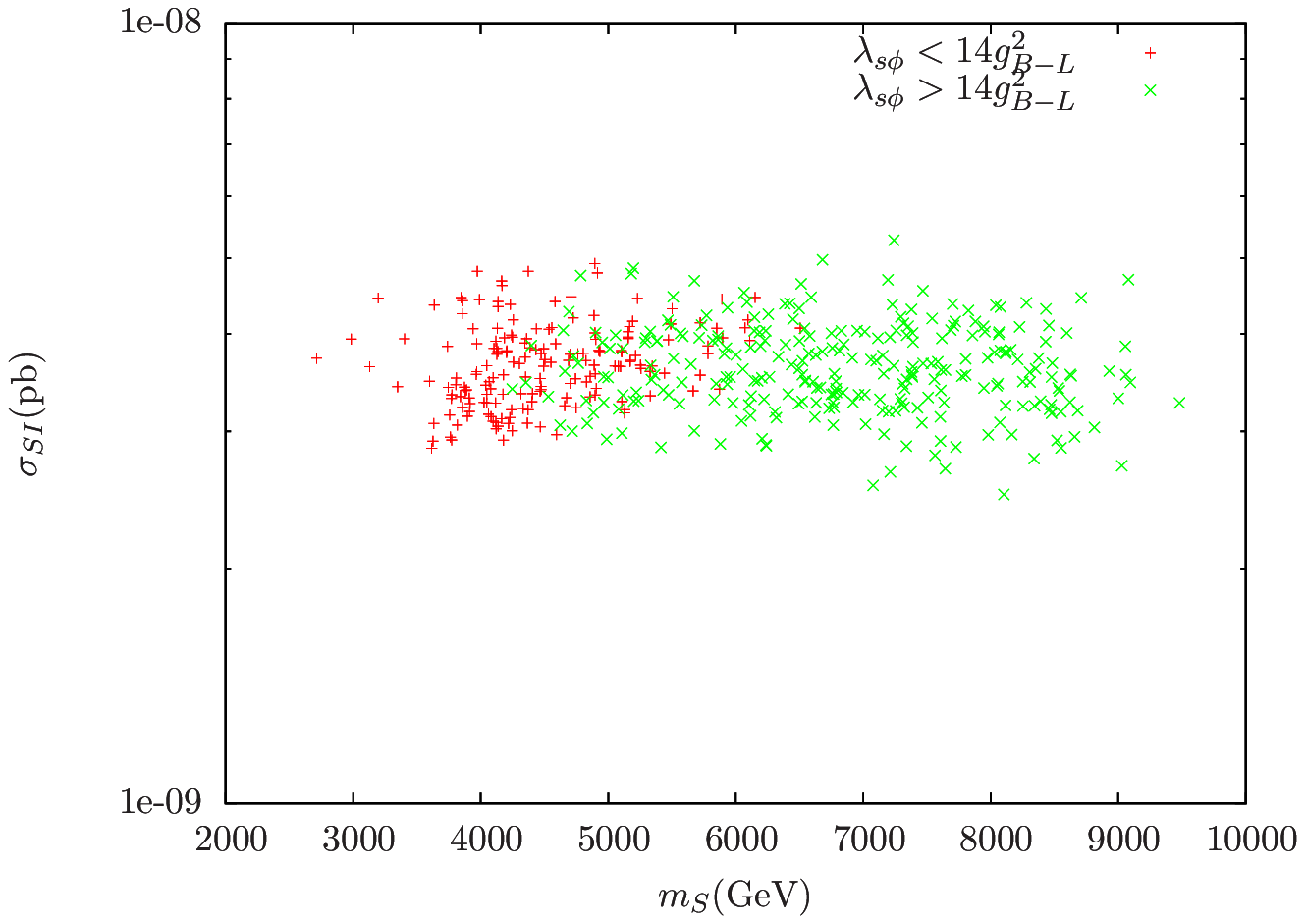}
\caption{Dark Higgs portal scenario on the $v_\phi-g_{B-L}$ plane. The LEP II excludes the region with $v_\phi<3.5$ TeV. The green and red points belong to the scenarios where DM and $Z'$ dominantly trigger CSI spontaneously breaking, respectively. We have imposed a perturbativity upper bound $\ld_{s\phi}\leq6.0$ in numerical scanning.}\label{dHiggs}
\end{figure}

%But the evolution of $\ld_{s\phi}$ is more complicated in the presence of a large $g_{B-L}$. From Eq.~(\ref{ru:sphi}) we know that the beta function of $\ld_{s\phi}$ now mainly receives two competitive contributions:  
%\begin{align}
%\beta_{\ld_{s\phi}}\approx\frac{\lambda_{s\phi}}{16\pi^2}
%\left(4\lambda_{s\phi}
%-24g_{B-L}^2\right). 
 %\end{align}
%When we start from a large $g_{B-L}$ at a scale $\Ld$, the beta function $\beta_{\ld_{s\phi}}$ may experience sign changing during the RGE flow, from negative to positive. Let the switch point be $\Ld_{int}$ at which $\beta_{\ld_{s\phi}}=0$. As long as $\ld_{s\phi}(\Ld_{int})<4\pi$ holds, the RGE analysis is reliable. Accordingly, $\ld_{s\phi}$ increases as $\Ld\ra \Ld_{int}$ but decreases as $\Ld_{int}\ra$ TeV. In Fig.~\ref{gBL} we demonstrate this behavior. Loosely speaking, in this way one shifts the UV boundary from $\Ld$ to $\Ld_{int}$ (with respect to the running of $\ld_{s\phi}$ with $g_{B-L}\ra 0$). Therefore, starting from a small $\ld_{s\phi}$ at $\Ld$ we eventually arrive a relatively large $\ld_{s\phi} \lesssim O(5)$ at TeV scale. By contrast, in the case of a negligible $g_{B-L}$, at low energy we can only have  $\ld_{s\phi}$ smaller than 1, which is too small to account for the thermal cross section Eq.~(\ref{SS2phi}), in particular for a larger $v_\phi$. For example, if $v_\phi=5$ TeV, $\ld_{s\phi}=6.4$ is needed. In summary, from Fig.~\ref{dHiggs} it is seen $g_{B-L}\gtrsim 0.3$, and thus $g_{B-L}$ should be large at $\Ld$. This means that the $\ld_{s\phi}(\rm TeV)\leq 6.0$ used before is reasonable. 

In the heavy DM scenario, an interesting picture may arise, namely, CSI spontaneously breaking can be triggered in practice 
mainly %implemented 
by DM field instead of $Z'$. This may provide a theoretical motivation for introducing a scalar DM field, 
and one of its immediate consequences is a prediction of a relatively heavy DM. We leave a more general 
discussion about such a scenario in the Section~\ref{con}. The presence of heavy scalar DM modifies 
the form of CSI spontaneously breaking. Now, the coefficients $B$ and $A$ given by Eq.~(\ref{CW:B}) and Eq.~(\ref{CW:A}) 
respectively receive new positive contributions from dark matter,
\begin{align}
\Delta B= +\f{1}{64\pi^2}\f{\ld_{s\phi}^2}{4}, \quad \Delta A= +\f{1}{64\pi^2}\f{\ld_{s\phi}^2}{4}\L-\f{3}{2}
+\ln \f{\ld_{s\phi}}{2}\R \,.
\end{align}
Note that for a complex DM scalar discussed later, there will be an extra factor 2. One can estimate 
when DM contribution to the CW potential becomes dominant over the $Z'$ contribution, Eq.~(\ref{CW:B}): 
\begin{align}\label{DM:SISB}
\ld_{s\phi}\gtrsim 2\sqrt{3}Q_\Phi^2g_{B-L}^2\simeq 14 g_{B-L}^2 \,.
\end{align}
In this scenario the PGSB mass is strongly correlated with the DM mass, with the ratio given by 
$m_\phi/m_{S}\approx \sqrt{\ld_{s\phi}}/4\pi$.  Therefore PGSB is typically much lighter than DM, given 
that $\ld_{s\phi}$ is constrained by perturbativity. Actually, because $\ld_{s\phi}$ now is forced to be 
$\sim O(1)$ by correct DM relic density, $m_\phi$ should be at least a few 100 GeVs. 
It is different from the $Z'$-dominant case where $m_\phi$ heavily depends on $g_{B-L}$ and thus can 
vary in a fairly wide region. This difference is manifested in the bottom panels in Fig.~\ref{spectrum}.

In Fig.~\ref{dHiggs} the green points are those satisfying the condition Eq.~(\ref{DM:SISB}). 
They are in the relatively large $v_\phi$ region ($\gtrsim3.5$ TeV), where a heavier $m_{Z'}$ can be 
realized with a relatively small $g_{Z'}$. Since we have imposed an upper bound $\ld_{s\phi}\leq6$, 
$v_\phi$ is bounded from above as  $v_\phi\lesssim6.5$ TeV. In our numerical study DM relic density 
within the region $\Omega_{\rm DM}h^2\in (0.09,\,0.12)$ is acceptable. Moreover, we only required that 
the branching ratio of DM annihilation into a pair of ${\phi}$ is larger than $50\%$ as the definition of dark 
Higgs portal scenario, and thus the actual $\ld_{s\phi}$ can be mildly beyond the region estimated 
by assuming the DM-nucleon scattering cross section in Eq.~(\ref{SS2phi}) to be $\sim$ 1 pb. 
Then the dark Higgs portal scenario is found out to works for $\ld_{s\phi}\in(1.0,\,6.0)$. 
Interestingly enough, if we take the perturbativity bound $g_{B-L}\lesssim 0.5$ (with $\Ld=M_{\rm Pl}$) 
seriously, the red points in the bottom panel of Fig.~\ref{dHiggs} are almost excluded. In other words, in the dark Higgs portal scenario DM as a trigger of CSI spontaneously breaking is favored by the perturbativity condition.

Now let us turn our attention to the DM direct detection rate in this scenario, where both $h_{\rm SM}$ 
and $\phi$ can give a sizable contribution to the DM-nucleon scattering. From Eq.~(\ref{DM:H}) it is noticed 
that the couplings for the $S-S-h_{\rm SM}$  and $S-S-{\phi}$ vertices are enhanced by $\ld_{s\phi}$ and 
$v_\phi$, respectively, and the suppression from the Higgs mixing angle can be compensated by these 
factors.   More explicitly, we have  
\begin{align}\label{SI:S}
\sigma_{\rm SI}=\L1-\f{m_h^2}{m_\phi^2}\R^2\times\frac{\lambda_{s\phi}^2}{4\pi}\f{ \mu^2 m_n^2}{ m_S^2m_h^4}\L f_N^{(n)}\R^2 \,,
\end{align}
in the limit $\ld_{sh}\ll \ld_{s\phi} \,$ so that the $h_{\rm SM}-$mediation is negligible. The prefactor can be traced back to the interference between two 
channels involving $h_{\rm SM}$ and ${\phi}$ exchanges in the $t$-channel. In particular, in the degenerate limit
$m_h\approx m_\phi$, the cross section vanishes. By contrast, if $m_\phi^2\ll m_h^2$, the cross section 
will instead scale as $\sim 1/m_\phi^4$, and thus enhanced. But in the dark Higgs portal scenario this would  
never happens, because of the ${\phi}$-mass distribution in the bottom right panel in Fig.~(\ref{spectrum}). Therefore we can neglect this factor and then the cross section is enhanced only by $\ld_{s\phi}^2/\ld_{sh}^2$, compared to that in the Higgs portal scenario, Eq.~(\ref{SI:H}). In the large $m_S$ limit,  the dark Higgs 
portal scenario is safe in the light of current LUX bound. Actually, it amounts to giving the bound: 
\begin{align}
{v_\phi}\gtrsim \sqrt{2\ld_{s\phi}}\times 1.0\times \L\f{10^{-8}\rm\,pb}{\sigma_{\rm SI,up}^{(n)}}\R^{1/2}\rm\,TeV,
\end{align}
which is always satisfied. But there is a good chance to observe or rule out the heavy DM in this scenario in the near future. As a matter of fact, from the lower panel of Fig.~\ref{dHiggs} 
we find that $\sigma_{\rm SI}$ is a few $10^{-9}$ pb which is within the sensitivity of the next generation 
of DM direct detection experiments. Besides,  that cross section is almost constant. The behavior is not surprising because $\sigma_{\rm SI}={\rm const}\times\L1-{m_h^2}/{m_\phi^2}\R^2\times  \sigma_{{\phi} {\phi}}v\,$, where the $ \sigma_{{\phi} {\phi}}v$ is the DM annihilation 
cross section that determines thermal DM relic density given in Eq.~(\ref{SS2phi}).  

Comments are in order. In the minimal analysis presented in this subsection, we have neglected three 
RNHs.  In principle it is possible that ${\phi}$ couples to RHNs with sizable couplings and hence the annihilating 
mode $SS\ra NN$ dominates over other modes. But the qualitative results in our previous analysis are not affected much, since the cross section for $SS\ra NN$ would scale as $\sim \ld_{N}^2m_S^2/v_\phi^4$, similarly to the previous case Eq.~(\ref{SS2phi}) except for the enhancement factor $\ld_N^2>1$, which however is offset  by a significant phase space suppression $\sqrt{1-M_N^2/m_S^2} \,$.  

%---------------------------------------------------------------------------------------------------------
\subsection{Modifications from the mixing effects}
%---------------------------------------------------------------------------------------------------------

In the previous two sections, concerning the DM pair annihilation, we have considered interactions in the 
individual sectors, respectively the SM and $U(1)_{B-L}$ sectors, ignoring the mixing between 
$h$ and $\phi$.  But the $h-\phi$ mixing could make some difference. Actually, we have displayed its 
importance in the DM direct detection. In this subsection, we will comment on their effects in DM pair 
annihilation. 

First of all, from Eq.~(\ref{DM:H}), let us note that the Higgs portal is modified by the trilinear coupling 
$S-S-h_{\rm SM}$ which receives a contribution from Higgs mixing. The effect can be absorbed in the redefinition of 
$\ld_{sh}\ra \ld_{sh}+\ld_{s\phi}$, but this operation produces an effect in the quartic coupling for $S-S-h_{\rm SM}-h_{\rm SM}$ 
vertex. In the light DM region with $m_S\lesssim 100$ GeV,  we have $\ld_{s\phi}\lesssim {\cal O}(10^{-3})$, 
which typically is much smaller than $\ld_{sh}$ and the resulting shift would be negligible. 
For the heavier DM, the shift may be appreciable and could lead to deviations from the standard Higgs 
portal scenario. For instance, one may arrange a cancellation between $\ld_{sh}$ and $\ld_{s\phi}$ such 
that the trilinear term vanishes and  only the quartic coupling survives. But in our numerical search we do 
not consider such subtlety. 

Secondly,  $\phi$ inherits couplings of $h_{\rm SM}$ and consequently it mediates DM annihilating into SM particles. For $\ld_{sh}\ll \ld_{s\phi}$, the two mediators, in the massless limit or in the degenerate limit between ${\phi}$ and $h_{\rm SM}$, show cancellation in DM annihilation by virtue of orthogonality of the rotation matrix from $(h,\phi)$ basis to the mass eigenstates. It is in analogy with Eq.~(\ref{SI:S}). Obviously, there are two possible ways to avoid this cancellation: 
\begin{itemize}
\item The first way is to assume $m_\phi\gg m_S$ so that the ${\phi}-$mediation is suppressed. Then only 
the $h_{\rm SM}-$mediation would survive. Nevertheless, such a Higgs portal-like scenario is distinct from 
the conventional Higgs portal scenario, because for $\ld_{sh}\ll \ld_{s\phi}$ we do not have the contact 
interaction $S^2h_{\rm SM}^2$, which is crucial for $SS\ra h_{\rm SM}h_{\rm SM}$ to catch up with the 
mode $SS\ra VV$ in the limit $m_S\gg m_V$. In other words, this scenario does not respect   
the equivalence theorem. 
More explicitly, in the limit of massless final states, we have
\begin{align}\label{DM:VV1}
  {\sigma}_{VV}v\simeq \delta_V\f{1}{64\pi}\f{(\ld_{s\phi}+\ld_{sh})^2}{m_S^2} \,,
\end{align} 
while $ {\sigma}_{hh}v$ is suppressed given $\ld_{sh}\ll \ld_{s\phi}$ and $m_S^2\gg m_h^2$. 
Here the parameter $\delta_V$ is given by  $\delta_W=2$ and $\delta_Z=1$ respectively.
Note that the $VV$ mode is enhanced by the annihilation into the longitudinal component of $V$, which 
produces a factor $(2m_S^2/m_V^2)^2$~\cite{Djouadi:2005gi}. On the other hand, other modes, including 
into the SM Higgs boson pair, are suppressed by $1/m_S^4$ and thus are negligible. We rewrite 
Eq.~(\ref{DM:VV1}) in a more illuminating form:  
\begin{align}\label{GSB:eq}
 {\sigma}_{VV}v\simeq \delta_V\f{1}{32\pi}\f{\ld_{s\phi}} {v_\phi^2}=\delta_V {\sigma}_{{\phi} {\phi}}v.
\end{align}
with $  {\sigma}_{{\phi} {\phi}}v$ given in Eq.~(\ref{2phi}).  Interestingly, the Goldstone equivalence theorem seems to be recovered if we replace $h_{\rm SM}\ra {\phi}$. However, this is not true because we 
have dropped the ${\phi}-$mediation by assuming $m_\phi\gg m_S$ and thus practically $ {\sigma}_{{\phi} {\phi}}v=0$. If that assumption breaks, then the ${\phi}-$mediation will spoil the relation Eq.~(\ref{GSB:eq}).
\item By contrast, the second way is suppressing the $h_{\rm SM}-$mediation by letting $m_\phi$ near 
$2m_S$ so as to resonantly enhance the ${\phi}-$mediation. As a result, DM mainly pair annihilates into 
a pair of $W/Z$. The cross section can be calculated in terms of the formula in~\cite{Kang:2012bq} 
\begin{align}\label{DM:VV}
  {\sigma}_{VV}v\simeq \f{f_{R}}{m_S^4}\f{1}{m_S}\L\f{2m_S^2}{v_\phi}\R^2 \Gamma(h^*_\phi\ra VV)\approx\delta_V\f{f_R}{4\pi}\f{\ld_{s\phi}^2}{m_S^2}. 
\end{align} 
with $f_R\approx 1/(4-m_\phi^2/m_S^2)^2$ in the narrow width approximation. One can find the expression 
for $\Gamma(h^*_{\phi}\ra VV)$ similar to that in Ref.~\cite{Djouadi:2005gi}, taking mass of the virtual
$\phi$ equal to $2m_S$. We are  working in the limit $m_S^2\gg m_V^2$. As Eq.~(\ref{GSB:eq}), one can rewrite $ {\sigma}_{VV}v\approx32\delta_Vf_R {\sigma}_{{\phi} {\phi}}v$. Accordingly, when $m_\phi$ is not far from the resonant pole, e.g., $m_\phi=1.7m_S$ which gives $f_R\approx0.8$, the $WW$ channel will become totally dominant and can readily have a correct cross section for DM even for $\ld_{s\phi}\lesssim{\cal O}(0.1)$. As a matter of fact, $\ld_{s\phi}$ should be sufficiently small, otherwise $m_S$ is so heavy that ${\phi}$ is not able to approach the resonant pole. After all, ${\phi}$ is a PGSB. The DM-nucleon scattering rate is given by Eq.~(\ref{SI:S}). The resonant enhancement in DM annihilating means that the factor $\ld_{s\phi}/m_S^2$ can be small, and thus $\sigma_{\rm SI}$ may be more or less 
suppressed.
\end{itemize}

We do not make a separated numerical study on the mixing effects, but one can get an impression on them from Fig.~\ref{Higgsp}, where the scattered points are due to such effects. For illustration, we select the samples in the ${\phi}-$mediation case, i.e., the case $m_\phi$ near $2m_S$, and label them red.  As expected, they do have a relatively smaller $\sigma_{\rm SI}$.

%---------------------------------------------------------------------------------------------------------
\section{$U(1)_{B-L}$ charged complex aDM with accidental $Z_3$: Shining light on GeV scale }%---------------------------------------------------------------------------------------------------------

If the DM candidate is a $U(1)_{B-L}$ charged scalar, we can get a trivial aDM with an accidental $Z_2$ by assigning it a peculiar charge such that it can only couple to $\Phi$ via $|S_X|^2|\Phi|^2$, as discussed 
in detail  in the previous subsection. However, we can also consider a nontrivial case where the charge of 
$S_X$ is fixed by coupling to $\Phi$. By virtue of SI, the possible interiactions among $S_X$ and $\Phi$ would 
be $\Phi^3S_X$, $\Phi^2 S_X^2$,  $\Phi S_X |S_X|^2$ and $\Phi S_X^3$. The first term evidently renders 
$S_X$ unstable. The second and third operators (along with $|\Phi|^2\Phi S_X$ that is not listed) actually give identical $\Phi$ and $S_X$ up to a conjugate, so again $S_X$ is not stable. Therefore, $X=-2/3$ is the unique option and we get DM with an accidental $Z_3$ discrete symmetry~\cite{Z3:other}. It survives even after $U(1)_{B-L}$ spontaneously breaking~\footnote{We also regard it as a remnant subgroup of the local $U(1)_{B-L}$~\cite{Ko:2014nha}: $U(1)_{B-L}$  would break down to $Z_3$ due to the $\lambda_3$ term after $U(1)_{B-L}$ breaking from nonzero VEV  $\langle \Phi \rangle$. 
}. Now, the relevant interaction terms are given by
\begin{align}
-{\cal L}_{Z_3}=\ld_{s\phi}|S_X|^2|\Phi|^2+\ld_{sh}|S_X|^2|H|^2+\L\f{\ld_3}{3}\Phi S_X^3+c.c.\R+V(H,\Phi).
\end{align}
SI and EW symmetries spontaneously break in the same way discussed in Section~\ref{SB:spectra}, i.e., SI spontaneously breaks mainly at the $B-L$ sector by virtue of the heavy $Z_{B-L}$ and is then mediated to the SM sector via the Higgs portal term. In the above Lagrangian, the $\lambda_3$ cubic term of $S_X$ (the subscript will be dropped henceforth) does not contribute to the mass of $S$. So the mass of $S$ again is given by Eq.~(\ref{DM:mass}). 

The correct thermal relic density of complex scalar DM $S$ can be realized 
in this model due to newly open annihilation channels involving $U(1)_{B-L}$ gauge interactions and the 
$\ld_3-$term, in addition to the usual Higgs portal integration from $\lambda_{s\phi}$ term which is present in the real singlet scalar DM scenario too. 
Arguably, among the various contributions to DM pair annihilation, that from $U(1)_{B-L}$ gauge interactions 
are always subdominant to from dark Higgs portal, due to the heaviness of $Z'$ or/and smallness of $g_{B-L}$. 
In order to see this, let us estimate the order of magnitude of DM annihilating into the SM fermion pairs via
$Z'$ in the  $s-$channel,
\begin{align}\label{DM:ff}
 \sigma_{f\bar f} v\simeq \sum_{f}C_f\frac{4\pi\alpha_{B-L}^2 m_{S}^2 v_f^2 X^2}{(m_{Z'}^2-4m_{S}^2)^2} Q_{f}^2\ra \sum_fC_fQ_f^2 \f{v_f^2X^2}{4}\L\f{1}{32\pi }\f{\ld_{s\phi}}{v_\phi^2}\R,
\end{align}
where $v_f$ is the relative velocity between the DM pair in the CM frame when they freeze out. 
The summation is over all $U(1)_{B-L}$ charged fermions $f$ which are lighter than DM. 
For the parameters $C_f Q_f^2$, we have $C_lQ_l^2=1$, $C_NQ_N^2=4$ and $C_qQ_q^2=1/3$ 
for leptons, RHNs and quarks, respectively. Therefore, this cross section is merely about $10\%$ of 
$  \sigma_{\phi \phi}v$ which is given in Eq.~(\ref{SS2phi}). When $m_S>m_{Z'}$, the mode 
$SS^*\ra Z'Z'$ is kinematically allowed. But  its cross section is even smaller than the one given in 
Eq.~(\ref{DM:ff}), and this channel plays no important role.

So, we will concentrate on the $\lambda_3$ cubic term, which will give rise to a crucial difference in DM 
dynamics. It opens an effective annihilation channel which is characteristic for $Z_3$ models, i.e., the 
semi-annihilation $SS\ra S^*{\phi}$ (via the contact interaction)~\footnote{Actually, there is another process relevant for DM freezing out, namely the three to two annihilation mode $SSS\ra S^*S^*$, which decreases 
number density of DM also. But it is suppressed by phase space and thus is negligible except for a very large coupling of $|S|^4$.}.  The cross section for this channel is simply given by
\begin{align}
\sigma_{{\phi} S}v \simeq \f{1}{64\pi}\f{\ld_3^2}{m_S^2}\sqrt{\f{3}{4}-\f{\sqrt{r_\phi}}{2}-\f{r_\phi}{4}},
\end{align}
as usual $r_\phi=m_\phi^2/m_S^2$. It scales as $\ld_3^2/m_S^2$, 
rather than $\sim \ld_3^2/v_\phi^2$ which 
was the case for the previous channel. Therefore, for reasonably large $\ld_3$ and light ${\phi}$  in this 
dark Higgs portal scenario, the lighter DM can easily achieve correct relic density even in the absence of 
a large coupling. So, this scenario survives even for a boundary $\Ld=M_{\rm Pl}$. It is a distinct feature from the dark Higgs portal scenario considered in the accidental 
$Z_2$ case, where by contrast a quite heavy DM is needed. Obviously, the basic reason is ascribed to the separation of DM mass source from the main DM interactions for annihilating. 

We would like to add one comment: annihilation from the $\lambda_3$ cubic term is not related to 
DM-nucleon scattering at all, even if the Higgs mixing is significant. Of course, here DM possesses 
$\sigma_{\rm SI}$ like Eq.~(\ref{SI:S}) due to the term for mass source, $|S|^2|\Phi|^2$. But that expression 
should multiply a factor $1/4$, because we are considering complex scalar which is non-self conjugate. 
In addition, ${\phi}$ is favored to be light, so we replace $m_h\ra m_\phi$. 
Then the estimated spin-independent direct detection  cross section is given by
\begin{align}
\sigma_{\rm SI}\simeq 0.54\times 10^{-11}\L\f{m_S}{100\rm\,GeV}\R^2\L\f{5\rm\,TeV}{v_\phi}\R^4\L\f{50\rm\, GeV}{m_\phi}\R^4\rm pb ,
\end{align}
which is about two orders of magnitude below the current bound. It is quite sensitive to $v_\phi$.  We have 
taken a larger $v_\phi$, which is necessary to get a sufficiently light PGSB with the help of smaller $g_{B-L}$. 
This DM may leave its hints in the future detectors, especially for the relatively heavy DM near TeV.

\begin{figure}[htb]
\includegraphics[width=4.7in]{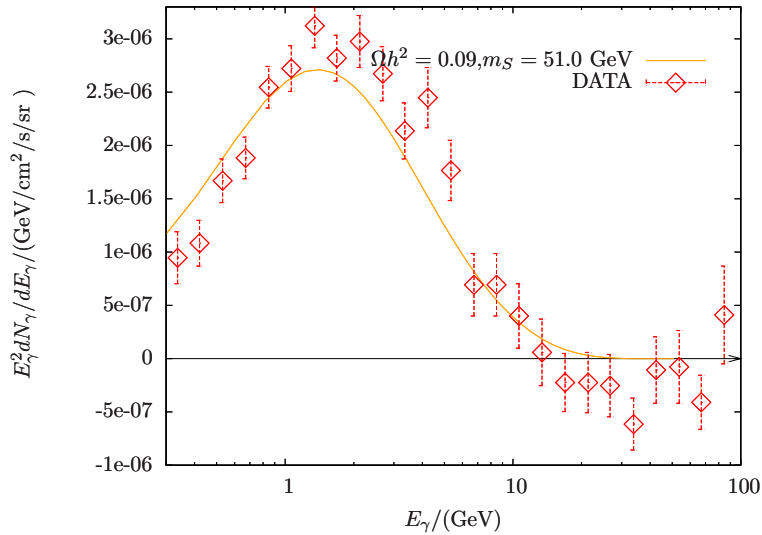}
\caption{The $\gamma-$ray spectrum of best fit point. In this fit we have taken the NFW dark matter density profile with $\gamma=1.20$. Data labeled as diamond stands for the residual spectrum used in our fitting. It depends on the diffuse background model, which suffers a large systematic uncertainty. Taking it into account leads to different residual spectrum, thus fairly different fit~\cite{Calore:2014xka}.}
\label{Z3}
\end{figure}

This $Z_3$ version of dark Higgs portal scenario is distinguished from other scenarios considered in this paper: it can accommodate the GeV $\gamma-$ray excess in the galaxy center while others can not. 
The excess was  claimed five years ago by L. Goodenough and D. Hooper after analyzing the Fermi-LAT 
satellite data~\cite{GeVex:1}. Its confidence has been increasing in the sequent years~\cite{GeVex,GeVex:2}. 
Light DM activities can account for the excess. For example, the signal is fit very well by a 31$-$40 GeV dark 
matter with $\sigma_{b\bar b}v = (1.4- 2.0) \times 10^{-26}$ cm$^3$/s~\cite{GeVex:2}. In this paper we have 
the Higgs portal scenario to accommodate such a DM. However, it has been excluded by DM direct detection, 
see Fig.~\ref{Higgsp}. Alternatively, one can adopt the hidden sector approach, where DM annihilates into two 
on-shell particles that subsequently decay into the SM fermions~\cite{Ko:2014loa,Boehm:2014bia}. Here we expect signature 
from the semi-annihilation $SS\ra S^* {\phi}$ followed by ${\phi}\ra b\bar b$, similar to the second scenario. 
In Fig.~\ref{Z3} we display the best fit, with $\chi^2=30$ that corresponds to $\chi^2/d.o.f.=1.25$, 
to the photon spectrum. It gives $m_S=51.0$ GeV,  
$m_\phi=44.3\rm\,GeV$ and a cross section that is slightly larger than the thermal one. The invisible Higgs branching ratio is negligible because the Higgs portal quartic coupling
is very small $\lambda_{h\phi} \lesssim 10^{-3}$. 
Actually, Ref.~\cite{Ko:2014loa} already includes this semi-annihilation mode in fitting the spectrum. 
But there it is merely one of the three annihilating modes and thus its role is not as important as in our paper, 
where the semi-annihilating mode is unique. 

%---------------------------------------------------------------------------------------------------------
\section{Discussion and conclusion}\label{con}
%---------------------------------------------------------------------------------------------------------

In this paper, we have explored the idea of accidental dark matter stability in the scale invariant local $U(1)_{B-L}$ model, which is also a theory for neutrino and at the same time radiatively breaks scale invariance via the $B-L$ sector dynamics. A real singlet scalar can be accidental  DM by an accidental $Z_2$, by virtue of both extended symmetries. A $U(1)_{B-L}$ charged complex scalar can also be a viable accidental DM due to an accidental (or remanant) $Z_3$. They can produce correct relic density via the annihilations from Higgs portal or dark Higgs portal, with the latter giving rise to an invisible DM at the direct detectors. The dark Higgs portal scenario is in tension with the LHC bound on $Z_{B-L}$, and only heavy DM of a few TeVs can have correct relic density. In particular, it may lead to DM triggering CSI spontaneously breaking. The situation is greatly relaxed in the $Z_3$ case which benefits the effective semi-annihilation mode and then light DM can be accommodated easily. Additionally, it is able to interpret the GeV gamma ray excess. This paper is based on MSIBL, but the main idea can be generalized to any local $U(1)_X$ (and nonabelian gauge group). Given the weakened constraints on $Z'$, the dark matter phenomenologies can be accommodated easily.

There are some open questions of interest. Firstly,  we mention that aDM from the higher $j\geq2$ multiplet has an interesting feature, i.e., it must be sufficiently heavy, at least a few TeVs, so as to suppress the too fast annihilation via the full EW interactions. Intriguingly, this is well consistent with the idea of DM triggering CSI spontaneously breaking, which also needs a heavy DM. Secondly, DM with a dark Higgs portal (or more general, dark portal) is present in many models. There dark matter, unlike the situation in this paper, may be probed only by means of indirect detections. However, the current bounds are far from clear. Therefore, it is worthy to employ an up-to-date analysis on this kind of models, with the current cosmic ray measurement data at hand. We leave answers to the open questions for further investigation.

\section{Acknowledgments}

We greatly thank Jinmian Li for providing us with the current LHC bound on $Z_{B-L}$. ZF would like to thank for the hospitality of Zhong Shan University, where a part of this work was done. This work was supported in part by  the Korea Neutrino Research Center which is established 
by the National Research Foundation of Korea(NRF) grant funded by the Korea government(MSIP) (No. 2009-0083526), and NRF Research Grant 2012R1A2A1A01006053 
(P.K. and Y.O.)

%---------------------------------------------------------------------------------------------------------
\appendix
%---------------------------------------------------------------------------------------------------------
\section{Relevant vertices}\label{coupling}
%---------------------------------------------------------------------------------------------------------
In this appendix we list the relevant vertexes that have been used in the text. Firstly, we present the coupling involving $\phi$, except for to dark matter:
\begin{align}\label{huh}
&\f{1}{2}\mu_{h\phi} h_{\rm SM}{\phi}^2~~ {\rm with}~~\mu_{h\phi}\approx\left[ \L\f{1}{2}-3\log 2 \R B\sin\theta-\ld_{h\phi} \L\cot\beta-2\sin\theta\R\right]  \,v_\phi,\cr
&\f{1}{2}\mu_{h\phi}' {\phi} h_{\rm SM}^2~~ {\rm with}~~\mu_{h\phi}'\approx \left[ \L\f{1}{2}-3\log 2 \R B\sin^2\theta -
 \right. \nonumber \\
 &\quad\quad\quad\quad\quad\quad\quad\quad\quad\quad\quad \left.
\ld_{h\phi} \L1+(2+3\ld_h/\ld_{h\phi})\cot\beta\sin\theta
-\f{7}{2}\sin^2\theta\R\right] \,v_\phi,\\
&y_{\phi f}{\phi} f\bar f~~ {\rm with}~~y_{\phi f}=-\f{m_f}{v}\sin\theta, \cr
& g_V{\phi} VV~~ {\rm with}~~ g_V=-C_V \f{m_V^2}{v^2}\sin\theta~ (\, C_W=2,\,\,C_Z=1).
\end{align}
At the leading order in $\sin\theta$, the trilinear coupling $\f{1}{3!}h_{\rm SM}^3$ is not affected and still given by $3\ld_h v$. Note that the couplings between two Higgs are derived taking into account the CW potential. Numerically, one can replace $B$ with $m_\phi^2/8v_\phi^2$. Compared to the $\ld_{h\phi}-$term, which is $\sim\sin^3\theta$, it can be much more significant in particular when $m_\phi\sim{\cal O}$(TeV). In that case we have $\mu_{h\phi}\approx -1.6Bv\lesssim{\cal O}$ (10) GeV. In some case it may plays a role. It is safe to make the approximation $\mu_{h\phi}'\approx -(1+3)\ld_{h\phi}v_\phi$ where the numerical factors 1 and 3 originate in the cross term $|H|^2|\Phi|^2$ and quartic term $|H|^4$, respectively. Next the trilinear couplings of DM to Higgs bosons are given by
\begin{align}\label{DM:H}
&\f{\mu_{sh}}{2}S^2h_{\rm SM},\quad \mu_{sh}=v\cos\theta(\ld_{sh}+\ld_{s\phi}\tan\beta \tan\theta)=v\cos\theta\L\ld_{sh}+\ld_{s\phi}\R,\cr 
&\f{\mu_{s\phi}}{2}S^2{\phi},\quad \mu_{s\phi}=v_\phi\cos\theta (-\ld_{sh}\cot\beta\tan\theta+\ld_{s\phi}).
\end{align}
It is safe to approximate $\mu_{s\phi}\approx \ld_{s\phi}v_\phi\cos\theta\approx 2m_S^2/v_\phi$, where the second approximation holds for DM getting mass mainly from coupling to $\Phi$, which is true for the heavier DM. As for the complex DM case, in the $\ld_{sh}\ra0$ limit we find the relevant interacting Lagrangian
\begin{align}\label{}
&-{\cal L}_{Z_3,int}=\left[\f{1}{3}\f{\ld_3}{\sqrt{2}}\L \cos\theta {\phi}+ \sin\theta h_{\rm SM}\R S^3_X +\f{\mu_3}{3}S_X^3+c.c.\right]+ \mu_{s\phi}|S_X|^2{\phi}+\mu_{sh}|S_X|^2h_{\rm SM},
\end{align}
with 
\begin{align}\label{}
\mu_3=\f{\ld_3}{\sqrt{2}}v_\phi,\quad \mu_{s\phi}=\ld_{s\phi}v_\phi \cos\theta,\quad  \mu_{sh}=\ld_{s\phi}v.
\end{align}

%---------------------------------------------------------------------------------------------------------
\section{Global minimum in a two scalar system}\label{global}
%---------------------------------------------------------------------------------------------------------

It is of importance to check that the vacuum considered in the text is the global minimum in the presence of an extra singlet scalar field $S$. If the vacuum with singlet developing VEV is the global minimum, DM will become unstable. This is a generic problem for the potential with multi-scalar. And we demonstrate this in a simplified model with two real singlets $S_1$ and $S_2$ respecting $Z_2$ symmetry, just the case in our paper. Their tree level potential is 
\begin{align}\label{}
V(S_1,S_2)=\f{\ld_1}{4!}S_1^4+\f{\ld_2}{4!}S_2^4+\f{\ld_{12}}{4}S_1^2S_2^2.
\end{align}
When one is interested in the case with only $S_1$ radiatively developing VEV $v_1$, triggered by a large $\ld_{12}$, the potential including radiative corrections can be rewritten as 
\begin{align}\label{}
V(S_1,S_2)=B_1S_1^4\L\ln \f{S_1^2}{v_1^2}-\f{1}{2}\R+\f{\ld_{12}}{4}S_1^2S_2^2+V_{\rm CW}(S_2)|_{Q=v_1}.
\end{align}
To derive it, we have eliminated $Q$ via the tadpole condition. The effective potential for $S_2$ is $V_{\rm CW}(S_2)|_{Q=v_1}=A_2 S_2^4+B_2S_2^4\ln (S_2^2/v_{1}^2)$. $A_{2}$ and $B_{1,2}$ are as usual functions of couplings only, and in the large $\ld_{12}$ limit they are given by
\begin{align}\label{}
B_{1}\approx B_2\approx \f{1}{64\pi^2}\f{\ld_{12}^2}{4},\quad A_2\approx    \f{\ld_2}{24}+\f{1}{64\pi^2}\f{\ld_{12}^2}{4}\L-\f{3}{2}+ \ln (\ld_{12}/2)\R. 
\end{align}

We now show that the desired vacuum with $v_1\neq0$ only is a global minimum with given conditions. Simply, its vacuum energy is $E_1=-B_1v_1^4/2$. Consider the other vacuum having $v_2\neq0$ only. Its value is related to $v_1$ via the tadpole condition again:
\begin{align}\label{}
v_2=v_1e^{-1/4-A_2/2B_2}.
\end{align}
Here $v_1$ appears due to the previous choice $Q=v_1$. The corresponding vacuum energy is $E_2=-B_2v_2^4/2$. Therefore, as long as $A_2/2B_2>-1/4$, we will have $v_2<v_1$ and hence $E_2>E_1$. 
For a relatively large quartic coupling $\ld_2$ such that $A_2/B_2>0$, this condition is always satisfied. In practice, the minimum of the second term of $A_2$ is about $-0.06$ for $\ld_{12}\approx 5.4$. Then $\ld_2\gtrsim0.14$ makes $A_2$ definitely positive. Additionally, one can easily see that the vacuum with both $S_1$ and $S_2$ developing VEVs is not a minimum for $\ld_{12}>0$. In conclusion, in this kind of two-scalar system one can choose proper parameters to guarantee the vacuum of interest is indeed the global minimum.

% largely speaking it amounts to show at other vacuum $v<v_\phi$. 

%---------------------------------------------------------------------------------------------------------
\section{RGEs}\label{RGE}
%---------------------------------------------------------------------------------------------------------

In this appendix we list the relevant RGEs for the gauge and quartic couplings. Moreover, we numerically study their evolutions from the low scale up to some high scale $\Ld$. Requring perturbativity if couplings at $\Ld$ has significant impact on the low energy phenomenologies. 
The RGEs for the gauge couplings are given by the following equations:
\begin{eqnarray}
\frac{dg_Y}{dt} & = & \frac{1}{16\pi^2}\frac{41}{6}g_Y^3, 
\\
%\end{eqnarray}
%
%\begin{eqnarray}
\frac{dg_{B-L}}{dt} & = & \frac{1}{16\pi^2}
\left[12g_{B-L}^3+\frac{32}{3}g_{B-L}^2g_{mix}+\frac{41}{6}g_{B-L}g_{mix}^2\right], 
\\
%\end{eqnarray}
%
%\begin{eqnarray}
\frac{dg_{mix}}{dt} & = & \frac{1}{16\pi^2}
\left[\frac{41}{6}g_{mix}\left(g_{mix}^2+2g_Y^2\right)
+\frac{32}{3}g_{B-L}\left(g_{mix}^2+g_Y^2\right)+12g_{B-L}^2g_{mix}\right].,
\end{eqnarray}
with $g_{mix}$ the kinetic mixing parameter between $U(1)_{B-L}$ and $U(1)_Y$. And it is assumed to be zero at the boundary. The RGEs for the quartic couplings are given by the following equations: 
\begin{eqnarray}
\frac{d\lambda_h}{dt} & = & \frac{1}{16\pi^2}\left[12\lambda_h^2+2\lambda_{h\phi}^2+\lambda_{sh}^2
-12Y_t^4+\frac{9}{4}g^4+\frac{3}{4}g_Y^2+\frac{3}{2}g^2g_Y^2+\frac{3}{2}g^2g_{mix}^2
+\frac{3}{2}g_Y^2g_{mix}^2+\frac{3}{4}g_{mix}^4 \right. \nonumber 
\\ 
& + & \left. \lambda_h\left(12Y_t^2-9g^2-3g_Y^2-3g_{mix}^2\right)\right], 
\\
%\end{eqnarray}
%
%\begin{eqnarray}
\frac{d\lambda_{h\phi}}{dt} & = & \frac{1}{16\pi^2} \left[
\lambda_{h\phi}\left(6\lambda_h+4\lambda-4\lambda_{h\phi}+6Y_t^2-\frac{9}{2}g^2
-\frac{3}{2}g_Y^2-\frac{3}{2}g_{mix}^2+Tr\left[\lambda_N^2\right]-24g_{B-L}^2\right)
\right. \nonumber 
\\ 
& - & \left.12g_{mix}^2g_{B-L}^2-\lambda_{sh}\lambda_{s\phi}\right], 
\\ 
%\end{eqnarray}
%
%\begin{eqnarray}
\frac{d\lambda}{dt} & = & \frac{1}{16\pi^2}\left[10\lambda^2+4\lambda_{h\phi}^2
+\lambda_{s\phi}^2-Tr\left[\lambda_N^4\right]+192g_{B-L}^4
+\lambda\left(2Tr\left[\lambda_N^2\right]-48g_{B-L}^2\right)\right], \label{ru:ld}
\\
%\end{eqnarray}
%
%\begin{eqnarray}
\frac{d\lambda_{sh}}{dt} & = & \frac{1}{16\pi^2}\left[
\lambda_{sh}\left(4\lambda_{sh}+6\lambda_h+6\lambda_s
+6Y_t^2-\frac{9}{2}g^2-\frac{3}{2}g_Y^2-\frac{3}{2}g_{mix}^2\right)
-2\lambda_{h\phi}\lambda_{s\phi}\right], 
\\
%\end{eqnarray}
%
%\begin{eqnarray}
\frac{d\lambda_{s\phi}}{dt} & = & \frac{1}{16\pi^2}\left[
\lambda_{s\phi}\left(6\lambda_s+4\lambda+4\lambda_{s\phi}
+Tr\left[\lambda_N^2\right]-24g_{B-L}^2\right)-4\lambda_{h\phi}\lambda_{sh}\right], \label{ru:sphi}
\\
%\end{eqnarray}
%
%\begin{eqnarray}
\frac{d\lambda_s}{dt} & = & \frac{1}{16\pi^2}\left[18\lambda_s^2+2\lambda_{sh}^2+\lambda_{s\phi}^2\right]. \label{lsRGE}
\end{eqnarray}

%FIG. 5 is the running of $\lambda_{s\phi}$. We took $\alpha_{B-L}=1$ at the Planck scale. 
%The quartic coupling $\lambda_{s\phi}$ is smaller than 5 at TeV scale.  

%---------------------------------------------------------------------------------------------------------

\end{document}